\documentclass[footinbib,twocolumn,amsmath,amstex,amssymb,mathfonts,superscriptaddress,prl]{revtex4}
\usepackage{graphicx}
\usepackage{color}
\usepackage{bm}

\usepackage{amsmath}
\usepackage{amssymb}
\usepackage{amsthm}
\usepackage{amsfonts}

\usepackage{bbm}

\begin{document}

\title{Topological dynamics of gyroscopic and Floquet lattices from Newton's laws}

\author{Ching Hua Lee}
\email{calvin-lee@ihpc.a-star.edu.sg}
\affiliation{Institute of High Performance Computing, A*STAR, Singapore, 138632.}
\affiliation{Department of Physics, National University of Singapore, Singapore, 117542.}
\author{Guangjie Li}
\author{Guliuxin Jin}
\author{Yuhan Liu}
\author{Xiao Zhang}
\email{zhangxiao@mail.sysu.edu.cn}
\affiliation{School of Physics, Sun Yat-sen University, Guangzhou 510275, China}
%\red{add yuhan's dispersive situation in the SOM as well as a counter example?}
%\blue{(Optional) Ask students to put in magnetic switching, Yuhan's counterexample etc in the appendix. No need to polish up. I will polish up and integrate them.}

\date{\today}
\begin{abstract}
%Of late, there has been intense interest in the realization of topological phases in very experimentally accessible classical systems like mechanical metamaterials and photonic crystals. Subjecting them to a time-dependent driving protocol further expands the diversity of possible topological behavior.
Despite intense interest in realizing topological phases across a variety of electronic, photonic and mechanical platforms, the detailed microscopic origin of topological behavior often remains elusive. To bridge this conceptual gap, we show how hallmarks of topological modes - boundary localization and chirality - emerge from Newton's laws in mechanical topological systems. We first construct a gyroscopic lattice with analytically solvable edge modes, and show how the Lorentz and spring restoring forces conspire to support very robust ``dangling bond'' boundary modes. The chirality and locality of these modes intuitively emerges from microscopic balancing of restoring forces and cyclotron tendencies. Next, we introduce the highlight of this work, a very experimentally realistic mechanical non-equilibrium (Floquet) Chern lattice driven by AC electromagnets. Through appropriate synchronization of the AC driving protocol, the Floquet lattice is ``pushed around'' by a rotating potential analogous to an object washed ashore by water waves. Besides hosting ``dangling bond'' chiral modes analogous to the gyroscopic boundary modes, our Floquet Chern lattice also supports peculiar half-period chiral modes with no static analog. With key parameters controlled electronically, our setup has the advantage of being dynamically tunable for applications involving arbitrary Floquet modulations. The physical intuition gleaned from our two prototypical topological systems are applicable not just to arbitrarily complicated mechanical systems, but also photonic and electrical topological setups.
\end{abstract}

\maketitle

Since topological concepts were first invoked to explain the curious formation of resistivity plateaus in quantum Hall systems\cite{tsui1982two,streda1982theory,arovas1984fractional}, they have lead to the discovery of novel condensed matter phases from topological insulators to Weyl semimetals\cite{qi2008,zhang2009topological,moore2010birth,qi2011,burkov2011weyl,zyuzin2012topological,xu2015discovery,soluyanov2015type,wang2016hourglass}. Various topological states are characterized by robust chiral boundary modes with potentially revolutionizing technological applications\cite{moore2009topological,hor2009p,garate2010inverse,zhang2010topological,zhang2012chiral,aguilar2012terahertz,liu2015switching,lee2015negative}, and have attracted substantial efforts in their search. Traditionally, the main focus had been on electronic topological materials, but the requisite precise bandstructure engineering\cite{haldane1988model,kane2005z,qi2006topological,tang2011high,sun2011nearly,neupert2011fractional,jian2013momentum,lee2014lattice} and sample control\cite{konig2007quantum,roth2009nonlocal} made realistic topological materials rather elusive. As such, there has been a recent paradigm shift towards more experimentally accessible realizations in optical lattices\cite{aidelsburger2013realization,miyake2013realizing,mancini2015observation,stuhl2015visualizing} as well as photonic\cite{wang2009observation,hafezi2011robust,yannopapas2012topological,kraus2012topological,hafezi2013imaging,mittal2014topologically,wu2015scheme,gao2015topological,lin2016dirac,lu2016symmetry,zhang2016topological}, electrical\cite{liang2013optical,pasek2014network,ningyuan2015time,PhysRevLett.114.173902,lee2017topolectrical,imhof2017topolectrical} and acoustic/mechanical\cite{kane2014topological,salerno2014dynamical,nash2015topological,wang2015topological,susstrunk2015observation,yang2015topological,zhu2015topologically,fleury2015floquet,lu2016observation} systems.

Amidst the plethora of topological systems, it is often unclear how topological properties exactly emerge from the physical equations of motion at the microscopic level. For instance, while quantum anomalous Hall (Chern) insulators have well-understood band topologies, it is hard to explain how their chiral Hall conductivity arises at the level of individual orbital behavior. Mere attribution to time-reversal symmetry breaking does not provide much illumination, since the system can still break time-reversal after a topological phase transition. Some limited intuition can be gained from studying the topological polarization\cite{fidkowski2010entanglement,yu2011equivalent,alexandradinata2011trace,lee2015}, which links charge pumping with the non-connectivity of the Hilbert space. Having intuitive understanding of topological behavior in terms of the fundamental equations of motion will not only be intellectually rewarding, but will also aids the design of new topological systems. 

The topological behavior of macroscopic mechanical lattices is particularly amenable to visualization, being ultimately governed by very intuitive Newton's laws. Mechanical topological systems also boast of unprecedented experimental accessibility and tunability\cite{paulose2015topological,susstrunk2016classification}, with analogs of quantum anomalous Hall and spin Hall states respectively demonstrated in experiments involving mechanical lattices of gyroscopes\cite{nash2015topological,wang2015topological,mitchell2016amorphous} and coupled pendula\cite{salerno2014dynamical,susstrunk2015observation}. The simplicity and ubiquity of such systems have expanded the realm of potential applications of topology to emerging technologies like heat diodes, adaptive materials, vibration isolation and acoustic waveguides\cite{paulose2015selective,ong2016transport,huber2016topological,liu2016topological}. 

The tunability of mechanical systems also opens up further topological possibilities through Floquet engineering. By dynamically varying one or more parameters, one may implement a time-periodic potential that extends the periodicity of Bloch states into the time domain. With an extra dimension in the configuration space, new topological invariants may be defined\cite{rudner2013anomalous,titum2016anomalous,lumer2016instability,potter2016classification}, leading to novel topological edge modes with no static analog\cite{kitagawa2010topological,rudner2013anomalous}. Indeed, chiral edge modes can emerge with appropriate Floquet modulation even if the undriven system is topologically trivial\cite{lindner2011floquet}. This has been demonstrated in the Haldane phase in Graphene realized through suitably designed irradiation\cite{oka2009photovoltaic,calvo2011tuning,kitagawa2011transport,morell2012radiation}, and is also purportedly useful for the realization of local and hence more realistic fractional Chern insulators (FCIs)\footnote{FCIs prefer flat bands, but the bands of local Hamiltonians cannot be flat unless they have vanishing Chern numbers\cite{chen2014impossibility,claassen2015position,lee2016band,read2016compactly}.}.  

As such, this work will focus on (i) providing very intuitive visualizations of how topological boundary localization and chirality can emerge from Newton's laws and, (ii) proposing a simple experimentally realistic set-up for the investigation and visualization of novel Floquet topological modes. 
To motivate intuition, we shall first introduce a mechanical realization of the simplest paradigmatic topological system, the Su-Schrieffer-Heeger (SSH) model. %\red{The intro is really nice, but we may want to add some about magnetic switching of gyroscope if we decide to talk it in the main text} 
It will then be naturally generalized to a 2D mechanical gyroscopic honeycomb lattice harboring chiral edge modes with analytic solutions. Building upon the particular robustness of this lattice, we next show how dynamically modulating it, without gyroscopes, can also give rise to chiral edge modes, some without static analogs. The emergence of topological behavior in these prototypical systems directly generalizes to generic systems described by Newton's or Maxwell's equations.
%Despite the simple forms Through  we can gain generic insights on the emergence of topological behavior.  
%We provide a rigorous study on the exact mechanism behind this driving protocol and, most significantly, provide a realistic experimental proposal for the realization of this topological Floquet mattress with electromagnets connected to AC currents.

\begin{figure}%[H]
\begin{minipage}{\linewidth}
\includegraphics[width=0.99\linewidth]{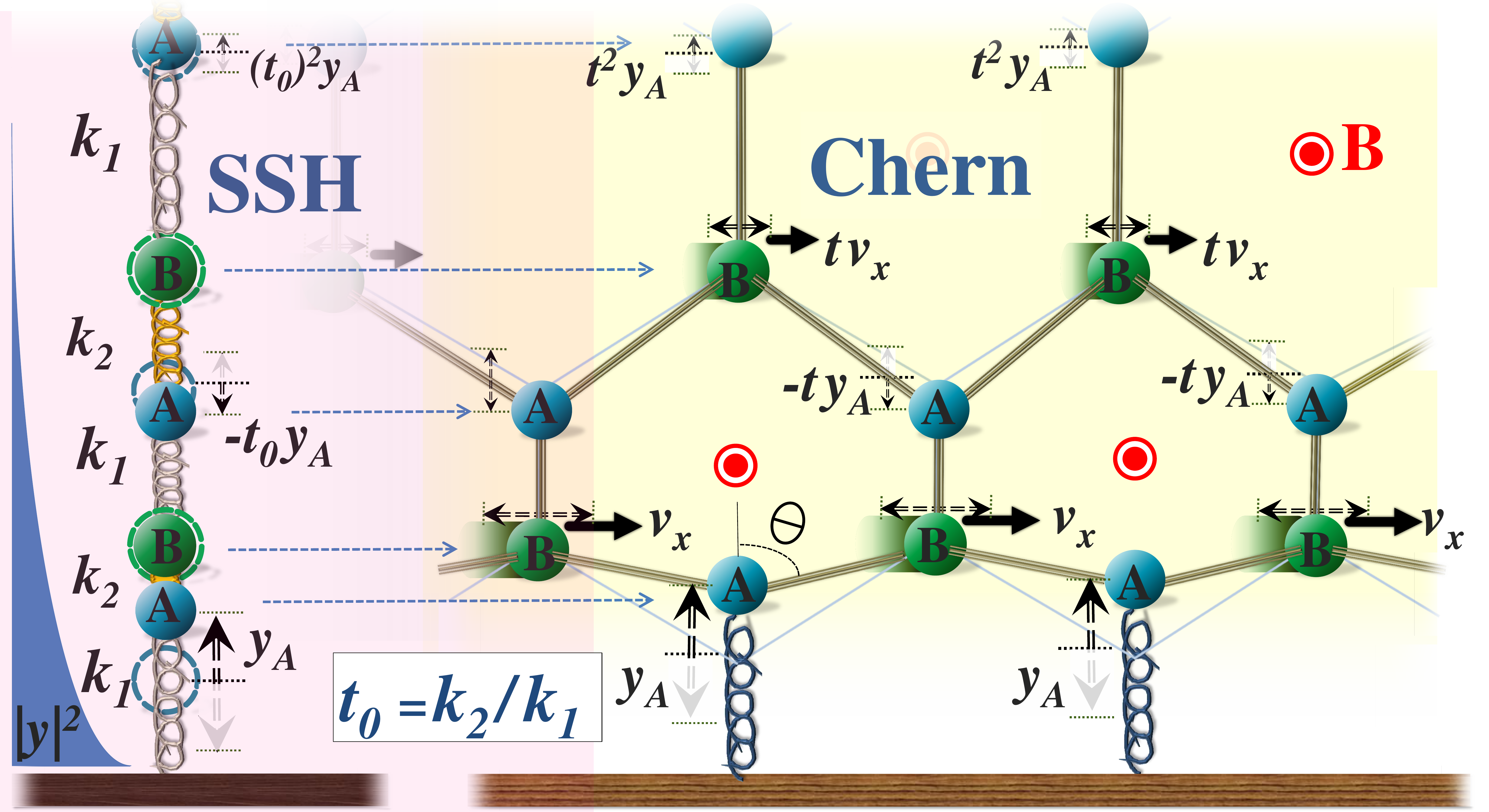}
\includegraphics[width=0.24\linewidth]{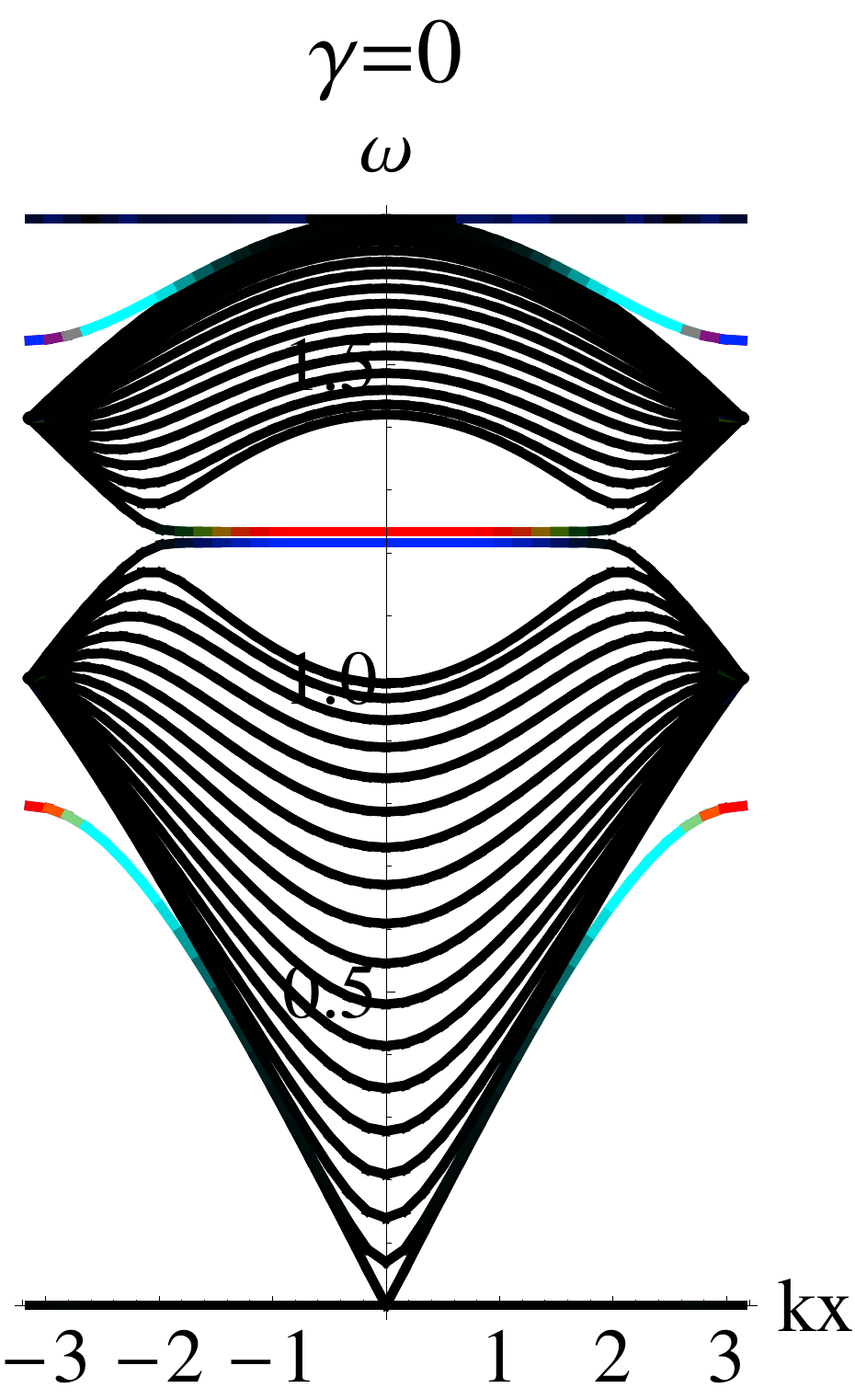}
\includegraphics[width=0.24\linewidth]{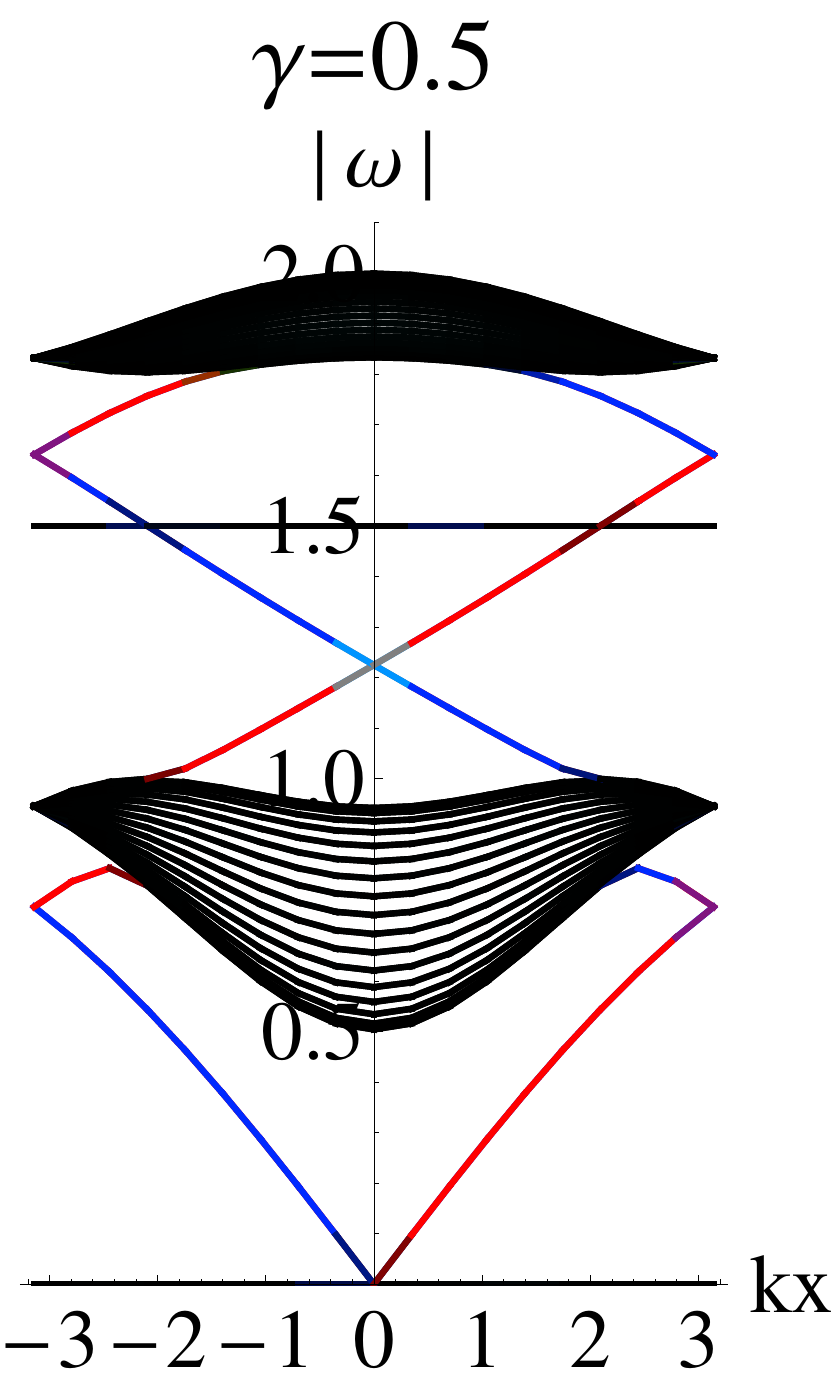}
\includegraphics[width=0.24\linewidth]{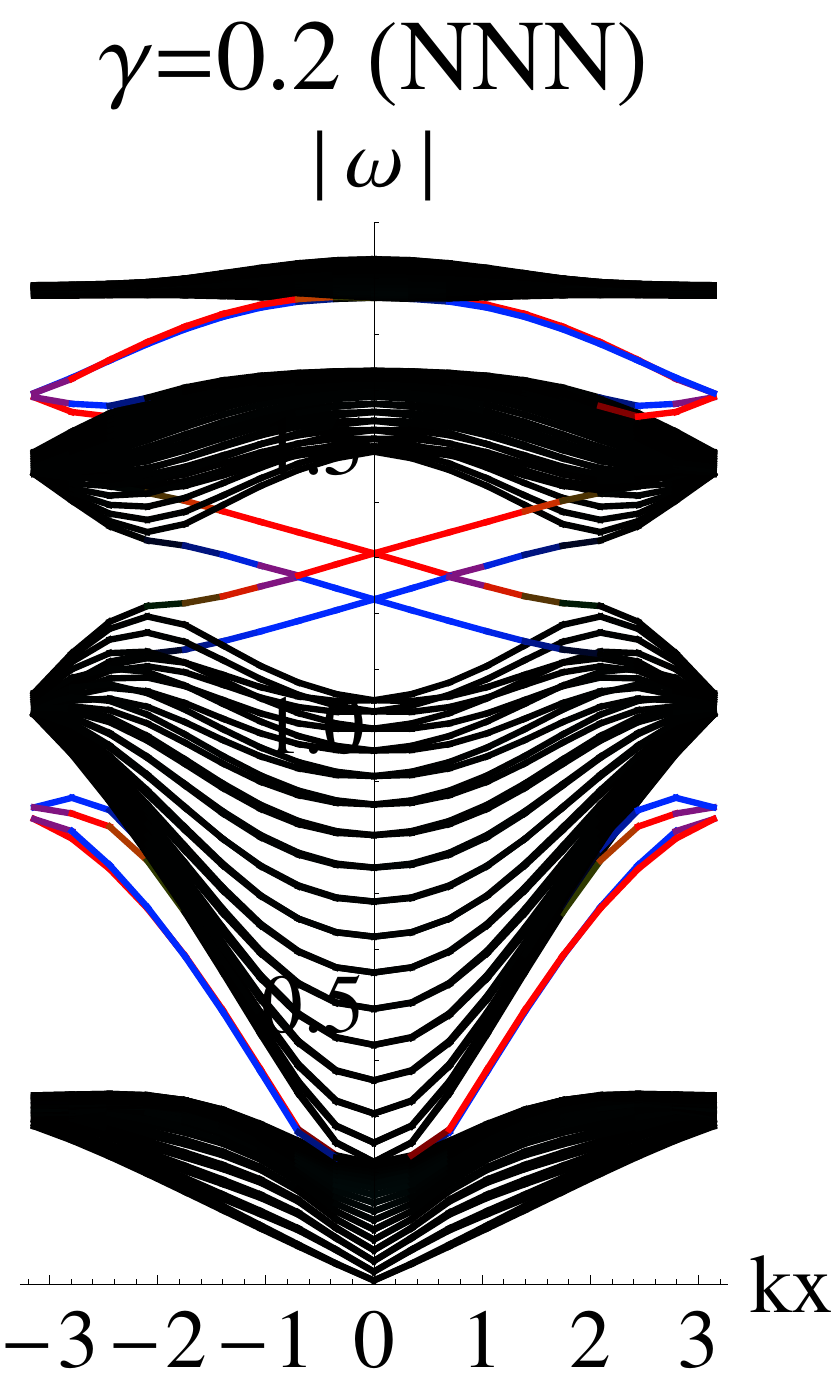} 
\includegraphics[width=0.24\linewidth]{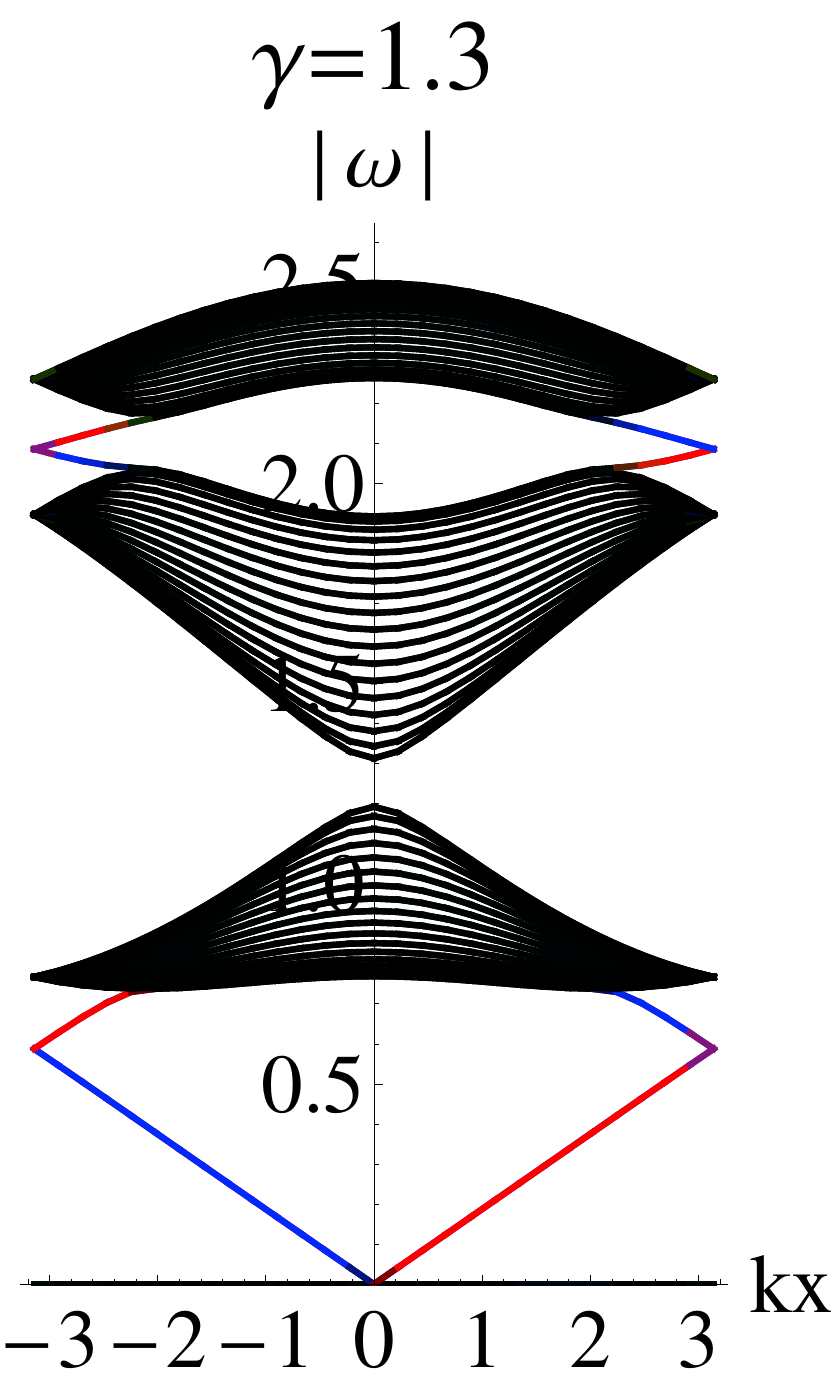}
\end{minipage}
\caption{(Color online) Top) A 2D Chern lattice (right) may be derived from a 1D SSH chain (left) by extending into additional perpendicular (horizontal) dimension, such that the vertical components of the spring constants remain the same. The SSH boundary mode consist of stationary green (B-type) masses and vertically oscillating blue (A-type) masses with exponentially decaying amplitude from the boundary. The Chern lattice inherits similar mechanical behavior, but with the green masses oscillating horizontally as well, to balance the magnetic Lorentz force. Bottom) Dispersion plots of a regular honeycomb lattice with NN identical springs $k=1$, masses $m=1$, and Lorentz/gyroscopic coupling $\gamma$. The intensity of the red/blue curves represent the extent of localization of the left/right edge modes. As $\gamma$ increase from $0$, the SSH-type boundary modes become chiral Chern edge modes with almost constant slope. The third figure with $\gamma=0.2$ was calculated with $~0.05$ NNN couplings that break the inversion symmetry, lifting the degeneracy of modes of opposite momenta. In the rightmost plot with $\gamma>\sqrt{3/2}$, the middle edge modes disappear as the bands undergo a phase transition from Chern numbers $0,0,0, 1$ to $-1,1,-1,1$, in ascending order. }
\label{fig:zigzag_large}
\end{figure}

{\it Mechanical SSH lattice--}
The simplest mechanical topological lattice is the mechanical analog of the SSH model, which ubiquitously described systems from polyacetylene to Majorana wires\cite{su1980soliton,takayama1980continuum,kivelson1981electron,fu2008superconducting,qi2008}. It consists of a semi-infinite chain of identical masses $m$ connected by springs of alternating stiffness $k_1$ and $k_2$ (Fig.\ref{fig:zigzag_large}). Suppose that $k_1>k_2$, where $k_1$ is the stiffness of the spring connected to a fixed boundary support. There exists an exponentially localized boundary mode with displacements of masses taking the form $\vec y=(y_A,0,-t_0y_A,0,t_0^2y_A,...)$, where $t_0=k_2/k_1<1$ and $y_A$ is the displacement of the first mass, which is connected to the boundary. This mode is an eigenmode because a stationary mass at position $2j$ remains stationary, experiencing zero net force from exactly balanced restoring forces of $k_1t_0^{j-1}y_A$ and $k_2t_0^jy_A$ from either spring. The ensuing oscillation of odd-numbered masses occur at frequency $\omega=\sqrt{\frac{k_1+k_2}{m}}$, which is a ``mid-gap'' mode well-separated from the continuum of bulk modes. 

This localized mid-gap mode can also be explained via its topological non-connectivity in momentum space. Key to this description is that, being a non-interacting topological phase, it does not require the non-commutativity of the density algebra, and can equivalently described in a configuration space whether classical or quantum. We first write down Newton's law as an eigenvalue equation: $-M \ddot \vec y =\omega^2\vec y= K\vec y$, where the mass-normalized stiffness matrix $M^{-1}K$ is the tight-binding ``Hamiltonian'', and $\vec y$ is the vector of displacements. In the basis of odd/even ($A/B$) sublattices, the (momentum space) $K$ matrix takes the form $\vec d(p)\cdot \vec \sigma=(k_2+k_1\cos p)\,\sigma_x+k_1\sin p \,\sigma_y$, where $\sigma_x,\sigma_y$ are the Pauli matrices. Due to inversion symmetry, there can be no $\sigma_z$ term and the eigenmodes define a mapping $S^1\rightarrow S^1$ from the periodic 1D Brillouin zone (BZ) to the circle $|\vec d(p)|^2=1$. The regime $t_0=k_2/k_1<1$ corresponds to a nonzero winding number of this mapping.  

One can generalize this 1D SSH system to a 2D mechanical lattice with analogously protected edge modes by adding an additional dimension such that the total vertical component of the spring stiffness between the $A$ and $B$ sublattices remain as $k_1$ and $k_2$ (Fig. \ref{fig:zigzag_large} Upper Panel). %\red{up panel is a good label for PRL? should we use a. b. to label}. 
For instance, a 2D honeycomb extension with identical springs $k$ and angle $2\theta$ between BAB masses (as illustrated) has $k_1=k$ and $k_2=k\cos^2\theta$, where $\theta=\pi/3$ for the rest of this paper. This 2D inversion symmetric system exhibit characteristic dispersionless edge modes inherited from the 1D SSH model, as shown in Fig. \ref{fig:zigzag_large} (Bottom Left). Just like their 1D counterparts, such edge modes consist of $\hat y$-polarized vibrations of $A$-type masses accompanied by a stationary $B$ sublattice.

{\it Mechanical chiral mode dynamics--}
The abovementioned 2D model can be driven into a Chern phase with chiral edge modes by breaking time-reversal symmetry. We first do so by introducing a gyroscopic (Lorentz) force $\vec F_B=\gamma\,\dot{\vec r}\times \hat z$, which 
 can arise either from a perpendicular magnetic field of strength $B=\gamma/Q$, where $Q$ is the charge on a mass, or from the reaction torque of a gyroscope attached to an oscillating mass. As detailed in our Supplementary online material (SOM) and Ref. \onlinecite{nash2015topological}, $\gamma=\frac{I\Psi}{h^2}$, where $I,\Psi$ and $h$ are the momentum of inertia, spin and length of a gyroscope respectively. Since magnetic forces are much weaker than electrostatic repulsion at nonrelativistic velocities, the gyroscopic interpretation is much more experimentally viable, and has indeed been successfully demonstrated\cite{nash2015topological}. %Nevertheless, we shall adhere to the equivalent but more intuitive magnetic field interpretation in the following discussions.
\begin{figure*}%[H]
\begin{minipage}{\linewidth}
\includegraphics[width=0.19\linewidth]{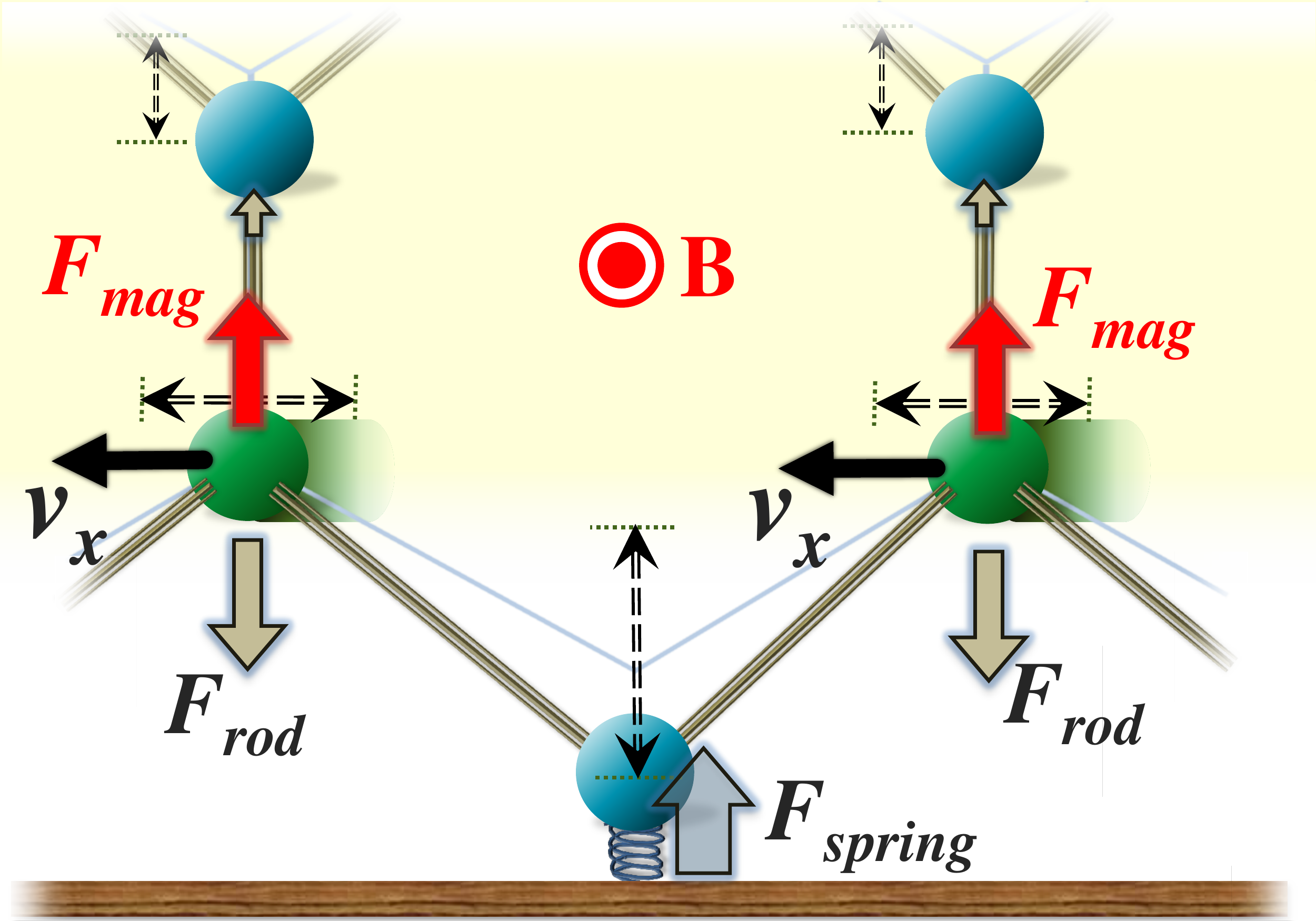}
\includegraphics[width=0.19\linewidth]{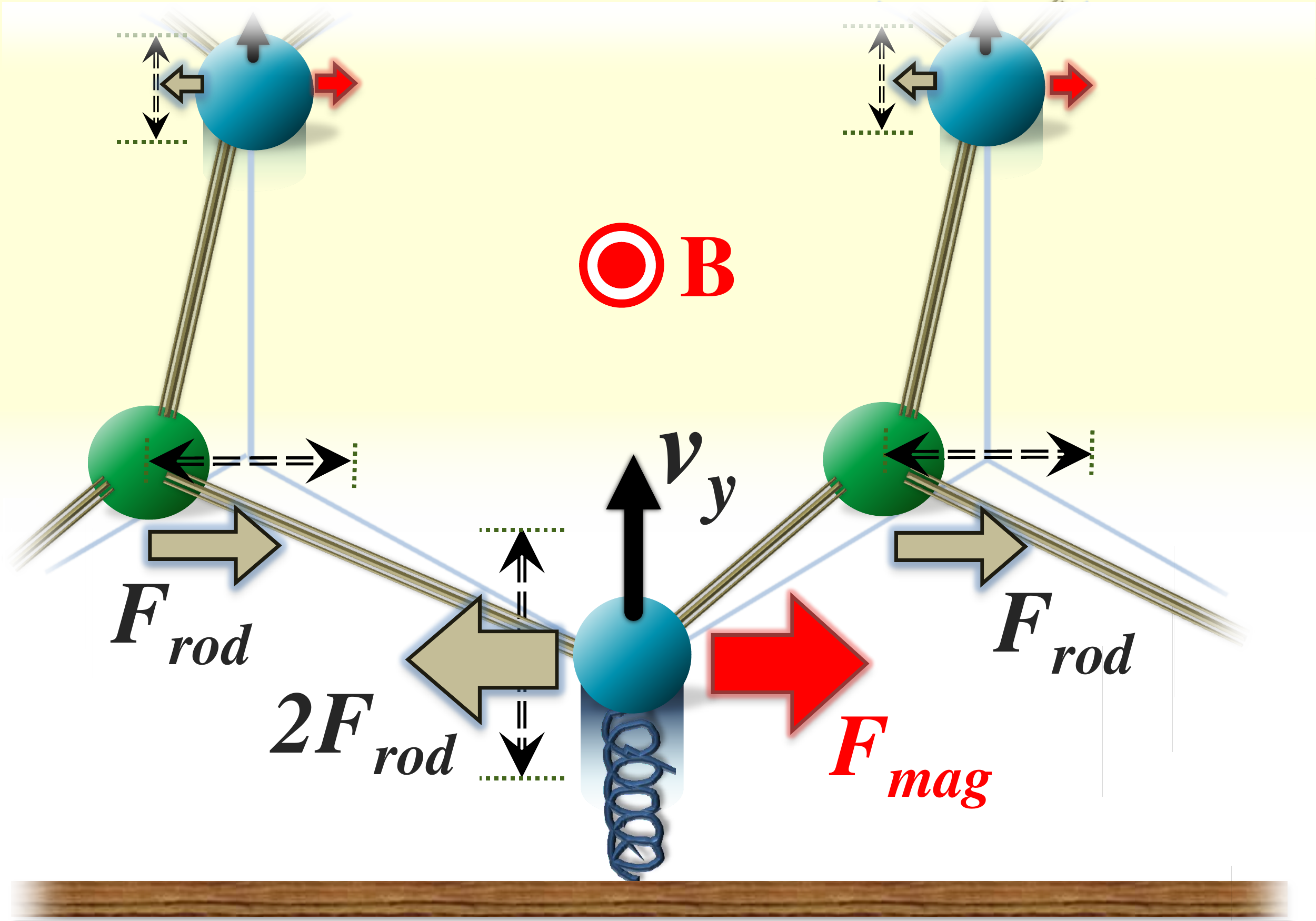}
\includegraphics[width=0.18\linewidth]{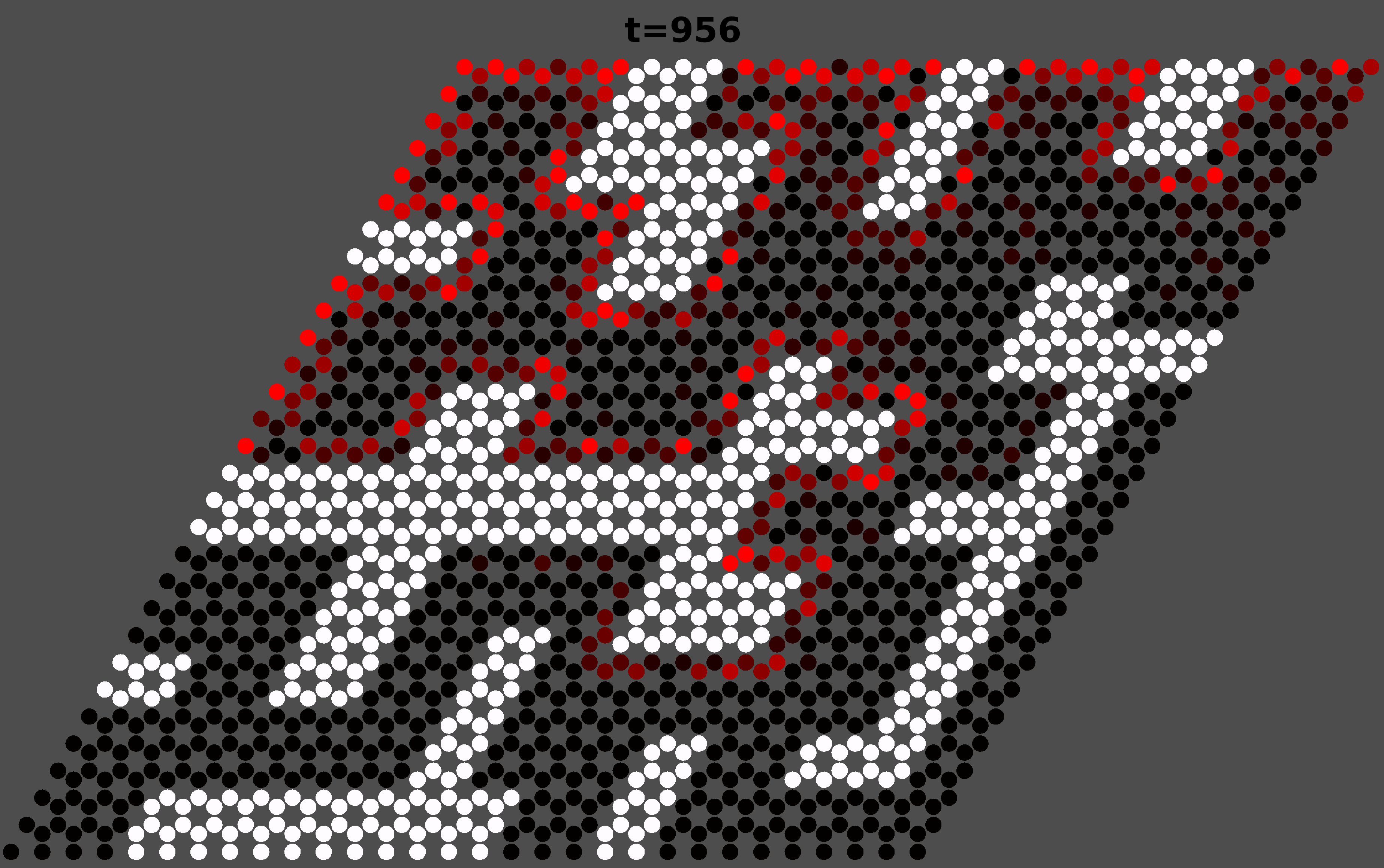}
\includegraphics[width=0.18\linewidth]{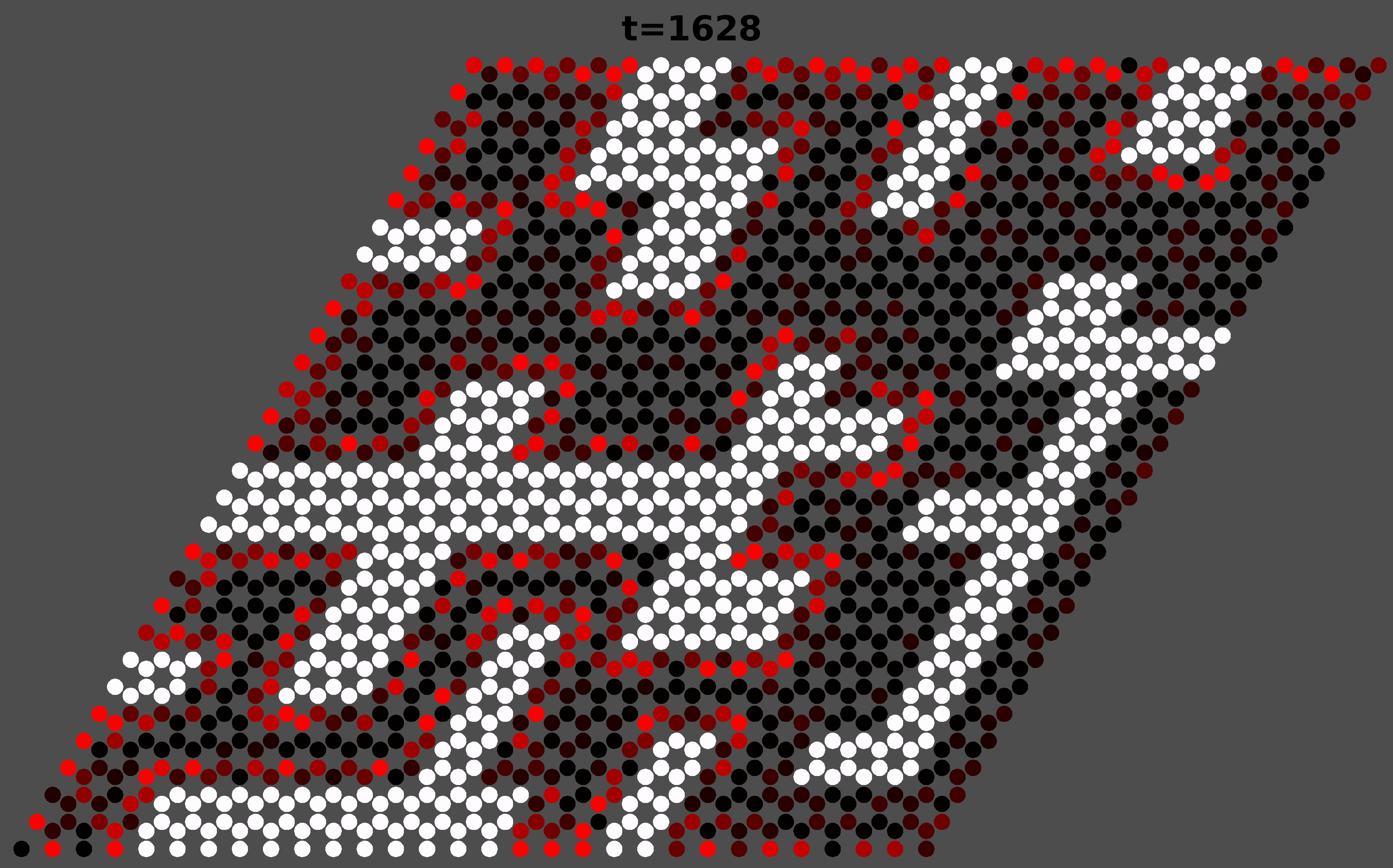}
\includegraphics[width=0.18\linewidth]{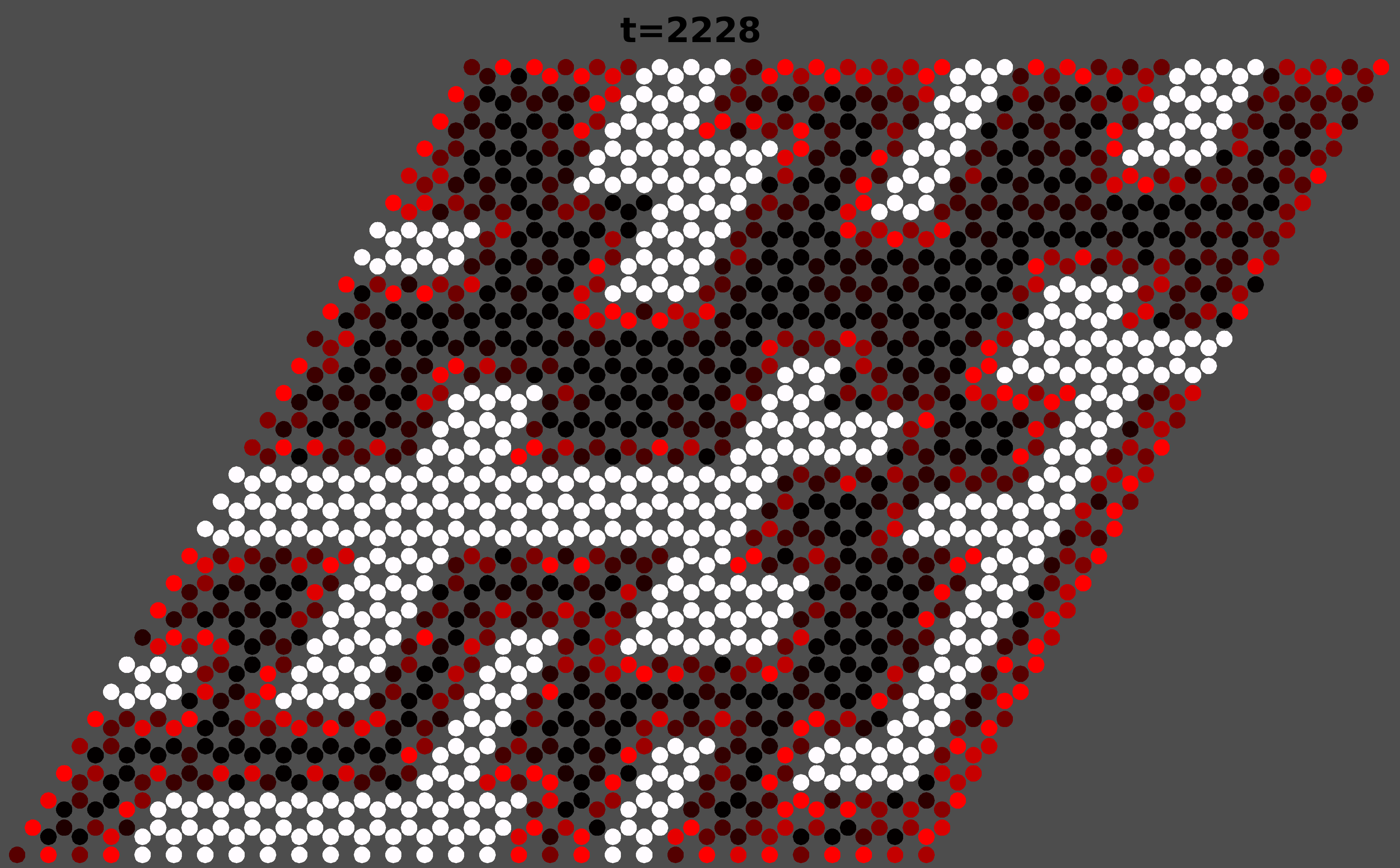}
\includegraphics[width=0.03\linewidth]{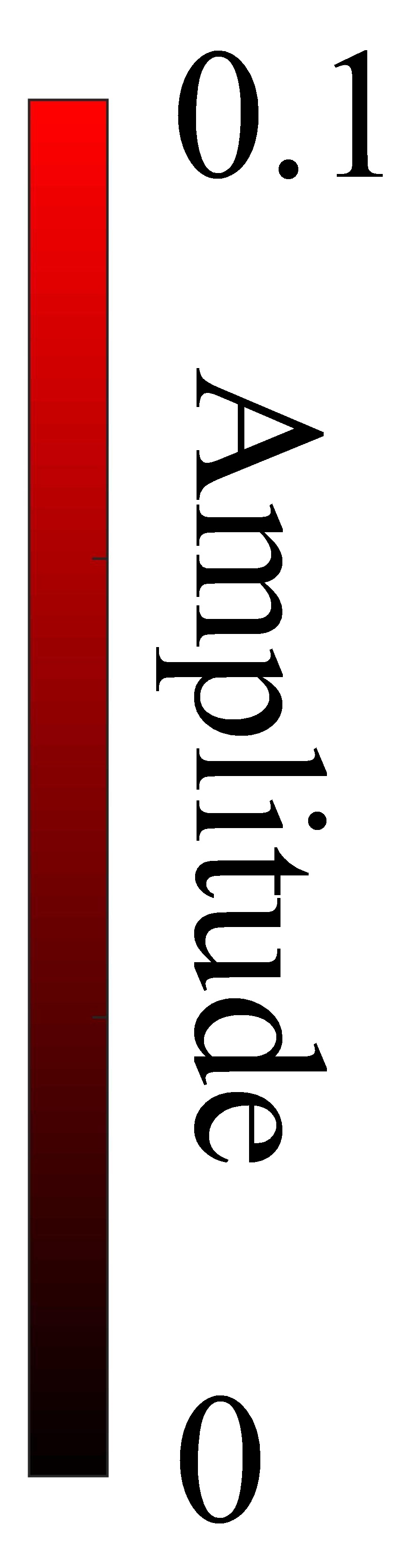}
\includegraphics[width=0.23\linewidth]{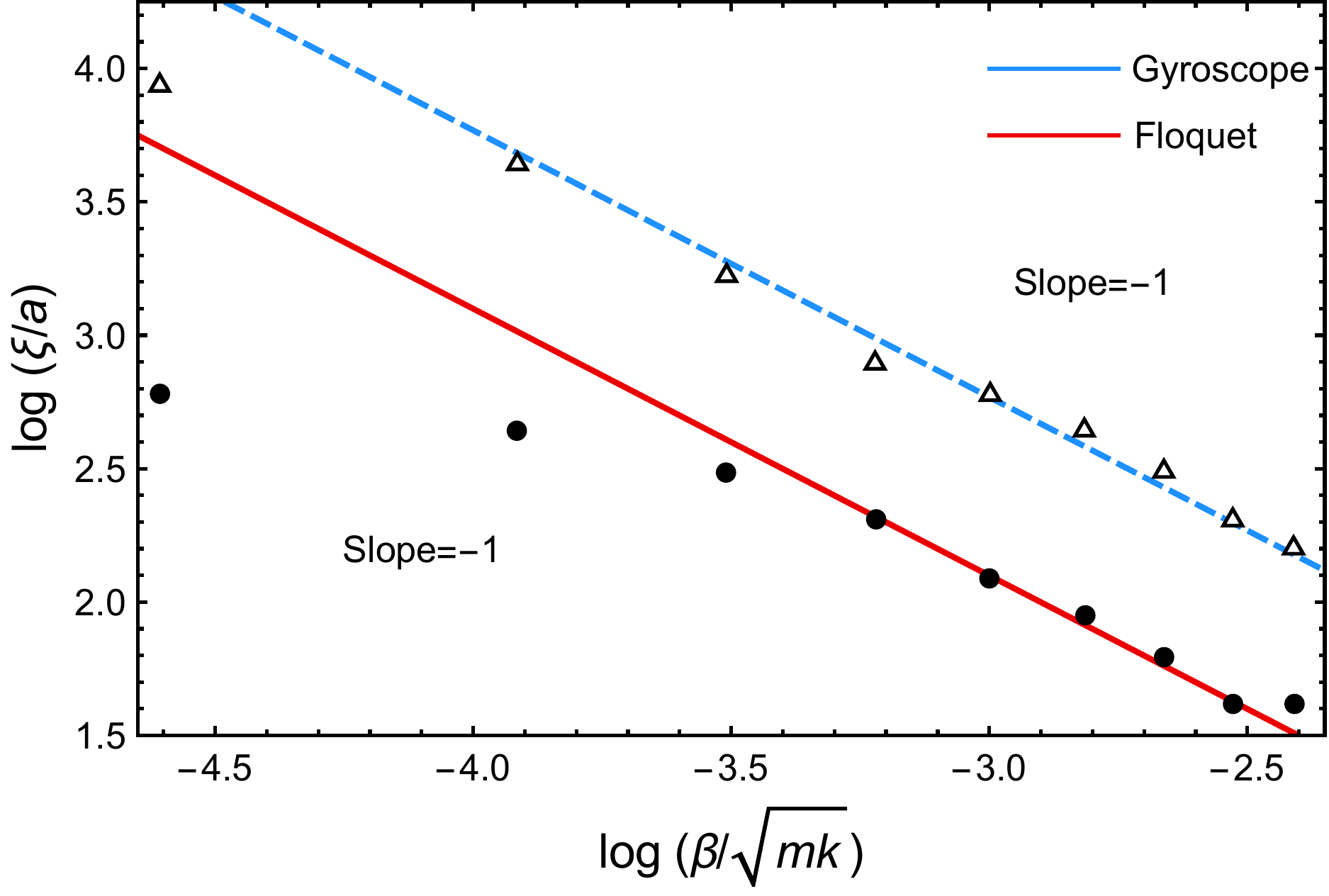}
\includegraphics[width=0.76\linewidth]{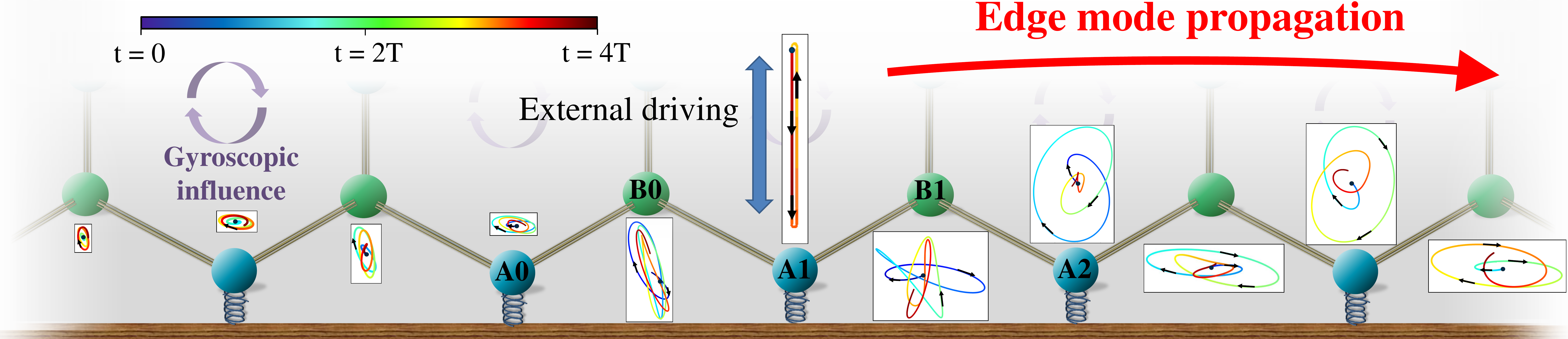}
\end{minipage}
\caption{( Top Left) Two snapshots of the chiral edge mode of the Chern lattice differing by a quarter cycle. On the left, the downward restoring forces on the green masses are exactly canceled by the Lorentz forces due to their motion, thereby leading to purely horizontal oscillations. This occurs analogously a quarter cycle later, though in the vertical direction with blue and green masses reversed. Top Right) Simulations (see SOM) depicting robust propagation of well-localized and minimally dispersive chiral edge modes on our honeycomb topological lattice, with white regions topologically trivial. Due to the near-perfect localization, the edge modes do not interfere even across fine features a few unit cells across. Bottom Left) Under significant damping $\beta$, edge modes propagate a distance $\xi\propto \beta^{-1}$ before extinction. Bottom Right) Simulation of boundary trajectories due to an externally excited mass A1. As described in the main text, the interplay between Lorentz forces, bulk restoring forces and dangling bonds results in an emergent chirality in mode propagation.}
\label{fig:maze}
\end{figure*}

With $\vec F_B$ included, Newton's 2nd law takes the form
\begin{equation}
M\ddot{\vec r}-i\gamma(\sigma_2\otimes \mathbb{I})\dot{\vec r}+K\vec r=0
\label{EOM}
\end{equation}
in the space spanned by the phonon polarizations and the sublattices. Eq. \ref{EOM} can be recast as an eigenvalue problem by rewriting it in the configuration space spanned by $U=(\dot{\vec r},\vec r)$. For each mode $\vec r$ at frequency $\omega$, we have
\begin{equation} 
\omega\,U=-i\left(\begin{matrix}
 & i M^{-1}\gamma \sigma_2\otimes \mathbb{I} & -M^{-1}K \\
 & 1 & 0\\
\end{matrix}\right)U=H_{\text{eff}}\,U
%(-M\omega^2-\gamma\omega\sigma_2\otimes \mathbb{I} +K)u=0...........
\label{EOM2}
\end{equation}
which is unitary equivalent to the time-dependent Schrodinger's equation considered in Ref. \onlinecite{zhang2010topologicalphonon}, but qualitatively distinct from that studied in Ref. \onlinecite{wang2015topological}, where the effects of the gyroscopes enter the mass term. %\red{why, is this the nice way to say paiwang's formula was wrong?} 
Due to the presence of both first and second order time derivatives, the configuration space for $H_{eff}$ has to be doubled. Indeed the effect of the Lorentz force/gyroscopic torque cannot be absorbed into an effective stiffness matrix, unlike the case of spin-orbit coupling in topological insulators, which enter as additional terms within the Hamiltonian of the \emph{first}-order quantum mechanical Schrodinger equation. Note that due to $PT$ symmetry, $H_{eff}$ has a real spectrum despite not being manifestly Hermitian. Had the gyroscopic term been replaced with a damping term $\beta(\mathbb{I}\otimes \mathbb{I})$, dissipation will have limited the propagation length of chiral edge modes (Fig. \ref{fig:maze} Bottom Left).   

Eq. \ref{EOM2} can be diagonalized to yield phonon dispersion spectra, shown in Fig. \ref{fig:zigzag_large} for the regular honeycomb lattice with identical springs $k$. Without a magnetic field ($\gamma=0$), it exhibits SSH-type dispersionless edge modes, with a similarly dispersionless bulk band at $\omega=\sqrt{3k/m}$ reminiscent of a Landau level with cyclotron orbits. With nonzero $\gamma$, the edge modes become chiral-propagating modes linking bulk bands with nonzero total Chern numbers. Landau level-like flat bands also appear at special values $\gamma=1/2$ and $\gamma=\sqrt{3}/2$. Given a bulk gap, we define $P$ to be the projection operator onto all bands below the gap. From well-known spectral flow arguments\cite{qi2011generic,alexandradinata2011trace,lee2013pseudopotential,rudner2013anomalous}, there exist $C=\frac1{2\pi i}\int Tr[P dP\wedge dP]$ edge modes traversing this gap. In our case with small $\gamma$, there are 8 bulk bands (only the 4 positive ones shown), with only the two bands of largest $|\omega|$ having Chern number $C=\pm 1$.

When extra next-nearest neighbor (NNN) hoppings are added to break the inversion symmetry, modes traveling in opposite directions are no longer degenerate. This is evident from the third dispersion plot with $k_{NNN}=0.05k_{NN}$ in Fig. \ref{fig:zigzag_large}, where a doubling of the edge modes (and less conspiciously, the bulk modes too) can be observed. %As contrasted with bandstructures from the first-order Schrodinger equation, there exist 

The masses in the edge modes behave in a particularly simple way. As illustrated in Figs.\ref{fig:zigzag_large} and \ref{fig:maze}, an uniform edge mode (at the $\Gamma$ point) consists of A-type (blue) ``dangling'' masses moving vertically, and B-type (green) masses moving horizontally. Although oscillations as such are attenuated in the bulk, they survive near the edge due to the interplay of the SSH mechanism and the Lorentz force. As detailed in the SOM~\cite{SOM}, elementary application of Newton's law on masses $A,B$ in both directions yields 
\begin{align}
\gamma\dot y_{A}+\frac{3k}{2}x_{B}&=0\\
m\ddot y_{A,n}&=-\frac{3k}{2}y_{A,n}\\
m\ddot x_{B}&=-\frac{3k}{2}x_B\\
\gamma \dot x_{B,n+1/2} &=k(1+t)y_A 
\label{EOMsim}
\end{align}
where $x_A$ and $y_B$ are the horizontal and vertical amplitudes of the oscillations of masses $A/B$, and $\sim t^{-x}$ the spatial profile of the edge mode. $t$ agrees with $t_0$ for the SSH system at zero Lorentz force coupling $\gamma=0$. From these equations, the edge mode is seen to exist due to the balance between the spring restoring forces and Lorentz forces: the dangling mass A moves only vertically because any horizontal force due to the magnetic field is canceled by the $\pi/2$ out-of-phase motion of mass B; reciprocally, mass B moves only horizontally because its Lorentz force and the spring restoring forces from all three A masses around it conspire to cancel. It is important to realize, however, that this seemingly intricate balance is actually topologically robust, existing continuously over a large range of spatial modulations wavenumbers $p_x$, as analyzed in the SOM~\cite{SOM}. 

This microscopic force analysis reveals the origin of the localization of the topological edge mode, which exists when $|t|<1$. From Eq. \ref{EOMsim}, it is equivalent to
%In this simple mechanical picture, the breakdown of the topological edge mode occurs when $|t|=1$, i.e. when an overwhelmingly large oscillation amplitude from within the bulk is required to maintain the balance of forces. $|t|$ can be easily solved from Eq. \ref{EOMsim} to be 
\begin{equation}
|t|=\frac{\gamma^2}{mk}-\frac1{2}<1,
\end{equation}
i.e. $\gamma<\sqrt{\frac{3mk}{2}}$. At $|t|=1$, the ratio of the amplitudes $\frac{x_B}{y_A}=\sqrt{\frac{2\gamma^2}{3mk}}=1$ (see SOM), and the edge mode breaks down since the oscillations are no longer large and forceful enough to compensate for the Lorentz force.%There the ratio of the amplitudes $\frac{x_B}{y_A}=\sqrt{\frac{2\gamma^2}{3mk}}=1$, which is manifestly at the limit of the compensatory ability of the Lorentz force. 

A more detailed analysis (see SOM) that takes into account spatial modulations yields
\begin{equation}
\omega(p_x)=\sqrt{\frac{3k}{2m}}-\frac{\sqrt{3}}{2}\frac{\gamma k}{3mk+2\gamma^2}p_x+O(\gamma^2)p_x^2
\end{equation}
with very small higher power corrections to the group velocity $-\frac{\sqrt{3}}{2}\frac{\gamma k}{3mk+2\gamma^2}$. This leads to very weakly dispersive edge modes that can travel robustly over long distances and across sharp bends. 
When $\gamma$ is further set to around the special value $\gamma=\sqrt{\frac{mk}{2}}$, $|t|$ vanishes for $p_x=0$ and the edge mode is perfectly localized at the edge, involving only the edgemost A and B masses. Such modes possess superior robustness in the presence of spatial disorder, and can circumnavigate complicated paths without disintegration (Fig. \ref{fig:maze}). 

Physical insight on the emergence of chirality can be gleaned from numerical simulation of the motion of individual masses in Fig. \ref{fig:maze} (Bottom Panel). Excitation of the center mass A1 causes all all nearby masses to oscillate. In particular, the two closest masses B0 and B1 accelerate in response to the displacement of A1, as well as to restoring forces from vertical springs connected to relatively stationary bulk masses. However, there is no reflection symmetry between the motions of B0 and B1 due to Lorentz forces, which causes both of them to curve clockwise. This in turn leads to differing amounts of energy propagation in either direction - in our hexagonal lattice, the trajectory of B0 is mostly orthogonal to the spring connecting it with A0, thereby suppressing motion towards the left. This suppression is especially pronounced due to the ``dangling'' nature of the A edge masses, each which can be excited only via their nearest B mass. In generic gyroscopic lattices, chiral edge modes will exist if the ``dangling'' edge masses are sufficiently isolated such that asymmetries in the motion of their very few neighbors possess overwhelming influence on their energy uptake.

{\it Realistic Floquet topological lattice--}
Interestingly, chiral edge modes can also exist in a lattice with time-dependent modulations, without invoking gyroscopic/Lorentz forces. We shall introduce a remarkably simple experimental proposal for such, building upon the honeycomb lattice. The combination of proximity to an SSH phase, minimal dispersiveness and excellent locality of edge modes make it particularly immune to imperfections.
\begin{figure}%[H]
\begin{minipage}{\linewidth}
\includegraphics[width=0.99\linewidth]{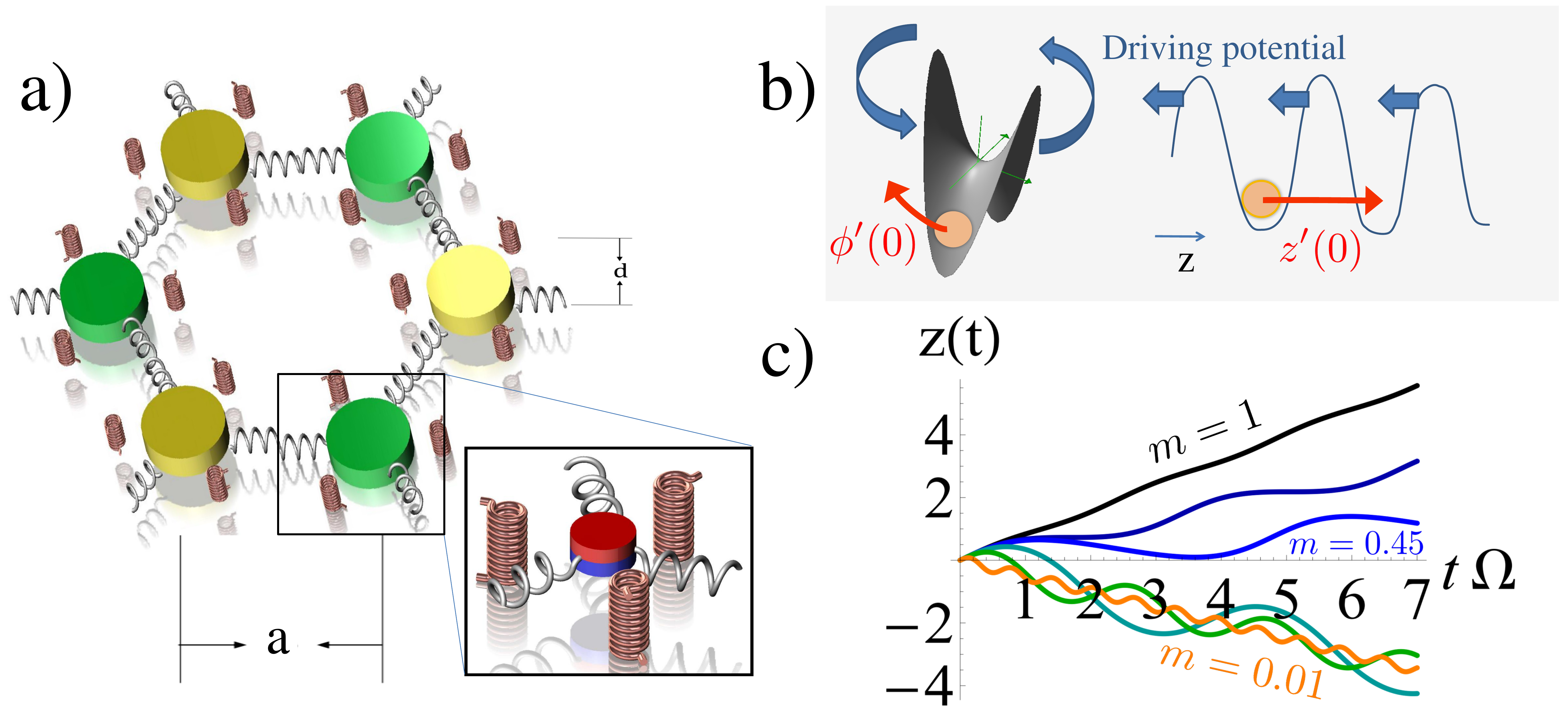}
\end{minipage}
\caption{(Color online)  a) Mobile permanent magnets (round masses) are surrounded by fixed AC electromagnets (coils), and mutually connected by springs in a honeycomb lattice fashion. b) Out-of-phase modulation of the electromagnets lead to an effectively rotating potential which, in the angular coordinate $z$, behave like traveling water wave. c) Exact solutions to the ``water wave'' motion (Eq.\ref{zddot}), with initial velocity $z'(0)=1$ \emph{against} the apparent motion of the potential. $m=1,0.5,0.45,0.25,0.1,0.01$ from top to bottom,  with all other parameters set to unity. A large mass ($m=1$) is  essentially unaffected, but smaller masses are yanked significantly by the potential. Very light masses ($m=0.01$) are overwhelmingly carried by the potential, and oscillate at the same period $\Omega$.
}
\label{fig:floquetexp}
\end{figure}

The set-up consists of a honeycomb lattice of mobile permanent magnets connected by springs $k$, such that each mobile magnet (mass) is surrounded by three fixed electromagnets at relative displacements $d\,\hat n_j=d(-\sin 2\pi j/3,\cos 2\pi j/3)^T$, $j=1,2,3$. (Fig \ref{fig:floquetexp}). When subject to AC currents, these electromagnets acquire time-dependent magnetic moments and hence fluctuating attractive or repulsive forces. A notion of chirality can be introduced by synchronizing the currents such that the moments of the fixed solenoids oscillate at a Floquet frequency $\Omega$ with relative phase offsets. Considering equally spaced offsets for simplicity, this adds a time-dependent part $K_t(t)$ to the stiffness matrix: 
\begin{eqnarray}
K_t(t)&=&  \Gamma \,\text{Re}\left[\sum_{j=1}^3 e^{2\pi ij/3}[\hat n_j\hat n_j^T]e^{i\Omega t}\right]\otimes\mathbb{I}\notag\\
&=&-\frac{3\Gamma }{4}\text{Im}\left[(\sigma_x+i\sigma_z)e^{i\Omega t}\right]\otimes\mathbb{I}
\label{Kt}
\end{eqnarray}
which corresponds to a potential $U_t(t)=\frac1{2}\vec u^TK_t(t)\vec u=-\frac{3\Gamma }{8}|\vec u|^2\cos(\Omega t+2\phi)$, $\phi$ the polar angle of the displacement $\vec u$. Here $\Gamma =\frac{3\mu_0 M_0 M_0'}{\pi d^5}$ in the limit of small oscillations, with $d$ the solenoid separation and $M_0,M_0'$ the magnetic moment amplitudes of the mobile magnet and fixed solenoids respectively. In other words, a particular displacement coordinate $\vec u$ is ``swept'' by a sinusoidal potential moving with frequency $\Omega$. This has the effect of nudging the mass in the direction of the sweep, since the mass generally receives more impulse when traveling in the direction of the sweep than vice versa. To see this more rigorously, we consider a simplified 1D version of this set-up, where $\phi$ is replaced by the coordinate $z(t)$ on a straight line:
\begin{equation}
m\ddot z(t) = - \nabla_z U_t(\Omega t + 2z(t))=-\frac{3\Gamma |\vec u|^2}{4}\sin(2z(t)+\Omega t)
\label{zddot}
\end{equation}
whose analytic solution (plotted in Fig.~\ref{fig:floquetexp}) is given by $z(t)=-\frac{\Omega t}{2}+J\left[\left(\frac{\Omega}{2}-z'(0)\right)t,\,A\right]$, where $J[w,A]$ is the Jacobi amplitude i.e. $\int^{J[w,A]}_0\frac{dt}{\sqrt{1-A^2\sin^2t}}=w$, and $A=\frac{12\Gamma |\vec u|^2}{m(\Omega-2z'(0))^2}$. Like an object among water waves, the mass is ``pushed'' by the potential when the latter is increasing and therefore, in its moving frame, experiences a potential with modified periodicity
%\begin{equation}
$T'=\frac{8\pi}{\Omega-2z'(0)}\frac{\text{K}\left[\frac{A}{A-1}\right]}{\sqrt{1-A}},$
%\end{equation}
with $J\left[K[A^2],A\right]=\frac{\pi}{2}$. This reduces to $T=\frac{2\pi}{\Omega}$ in the limit of large mass or small $A$, where the effective acceleration is small. It diverges at $\Omega=2z'(0)$, when the mass is able to follow the moving potential exactly. As summarized in Fig.~\ref{fig:floquetexp} b) and c), the dynamically driven potential drags/pushes the mass along akin to rotating water waves, hence introducing chirality in way qualitatively distinct from Lorentz forces.

To see that our periodically driven lattice indeed possess chiral edge modes, we employ the Floquet approach which gives the effective Hamiltonian for a out-of-equilibrium system averaged over one cycle. Consider a generic Hamiltonian with periodicity $T=\frac{2\pi}{\Omega}$, i.e. $H(t)=H(t+2\pi/\Omega)$. In analogy to Bloch's theorem, the eigensolutions to the Schrodinger's equation $(H(t)-i\partial_t)|\Phi(t)\rangle=0$ (Eq. \ref{EOM2}) can be decomposed as $|\Phi(t)\rangle=e^{i\omega t}\sum_m\phi_me^{im\Omega t}|m\rangle$, where $\phi_m$ are the Fourier components of the time-periodic part $|\Phi(t)\rangle$, and $\omega$ is its Floquet quasifrequency which lives in the ``energy'' BZ. Substituting this decomposition into the Schrodinger's equation and integrating over a period, one obtains the (time-independent) effective Floquet Hamiltonian~\cite{claassen2016all}
\begin{equation}
H_F=\sum_{mm'}\left(H_{m-m'}+m\Omega\delta_{mm'}\right)|m\rangle\langle m'|
\label{floquet}
\end{equation}
where $H_{m-m'}=\frac1{T}\int_0^TH(t)e^{i(m-m')\Omega t}dt$. % are the Fourier components of the original time-dependent Hamiltonian. \red{Hj is not defined?}
The Floquet quasifrequencies $\omega$ are the eigenvalues of $H_F$. Alternatively, Eq. \ref{floquet} may be viewed as a Wannier-Stark problem with $\Omega$ taking the role of a constant field strength in the extra ``Sambe space'' dimension spanned by $|m\rangle$, and Fourier terms $H_{m-m'\neq 0}$ causing jumps between different $|m\rangle$.

\begin{figure}%[H]
\begin{minipage}{\linewidth}
\includegraphics[width=0.25\linewidth]{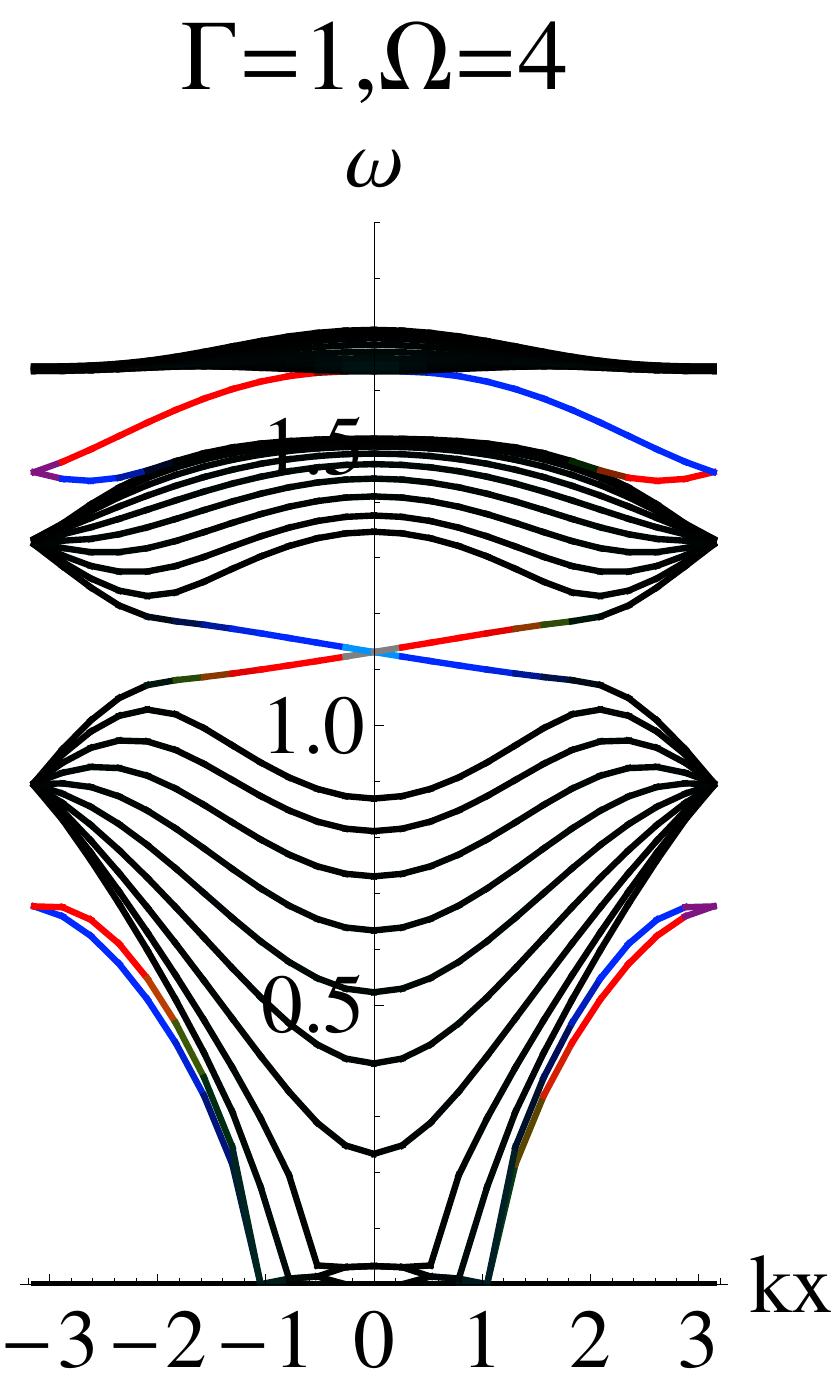}
\includegraphics[width=0.46\linewidth]{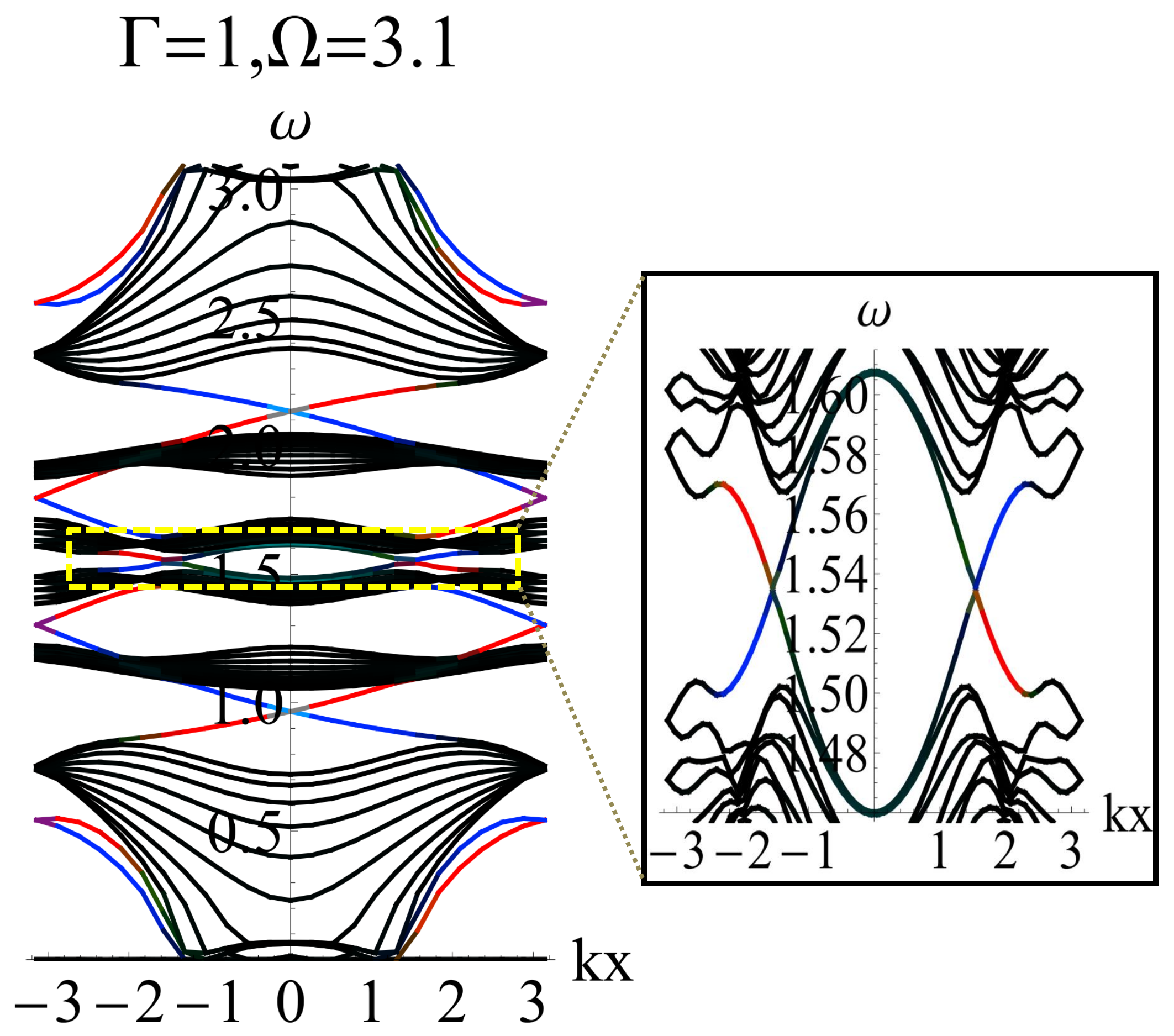}
\includegraphics[width=0.26\linewidth]{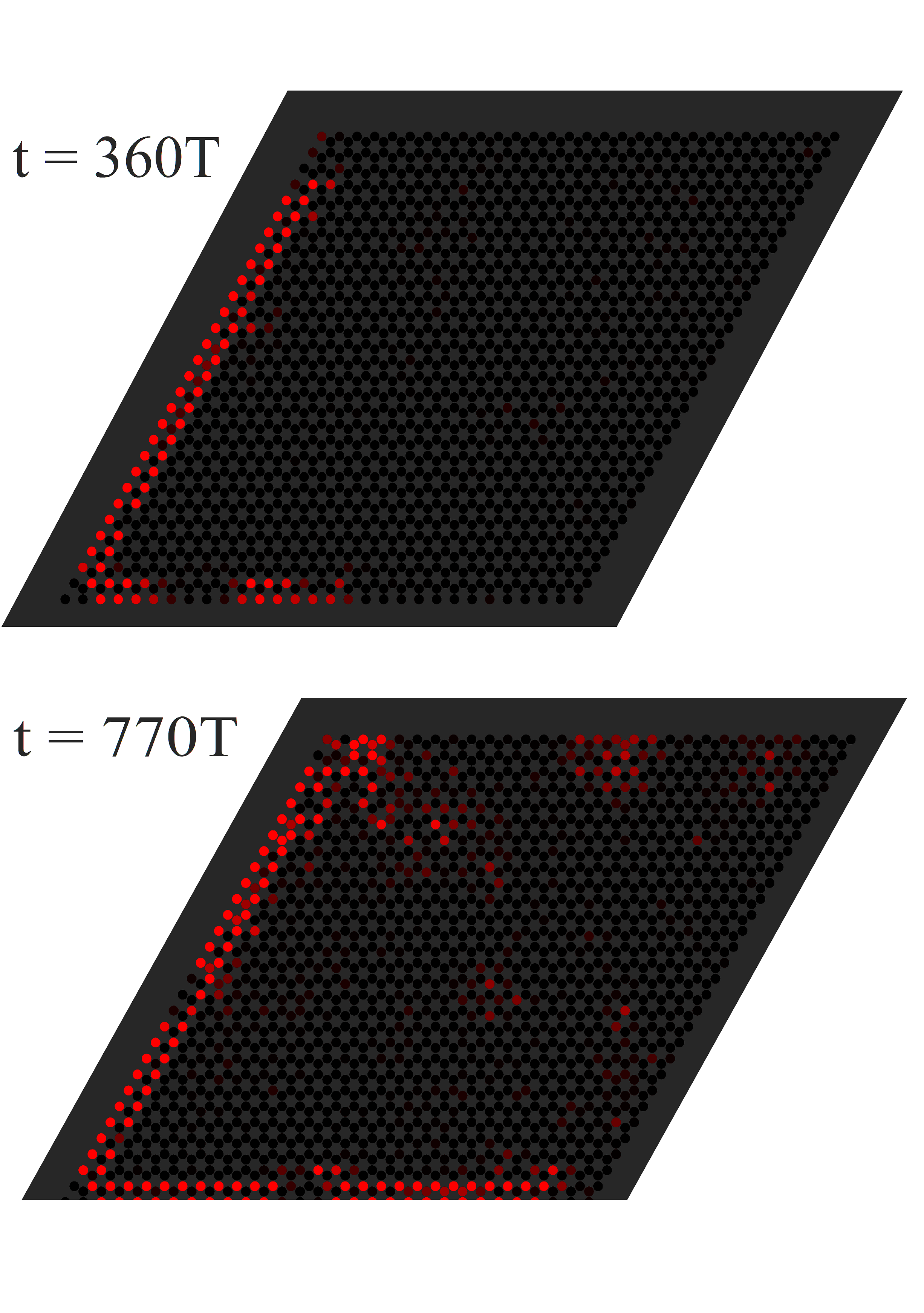}
\includegraphics[width=0.47\linewidth]{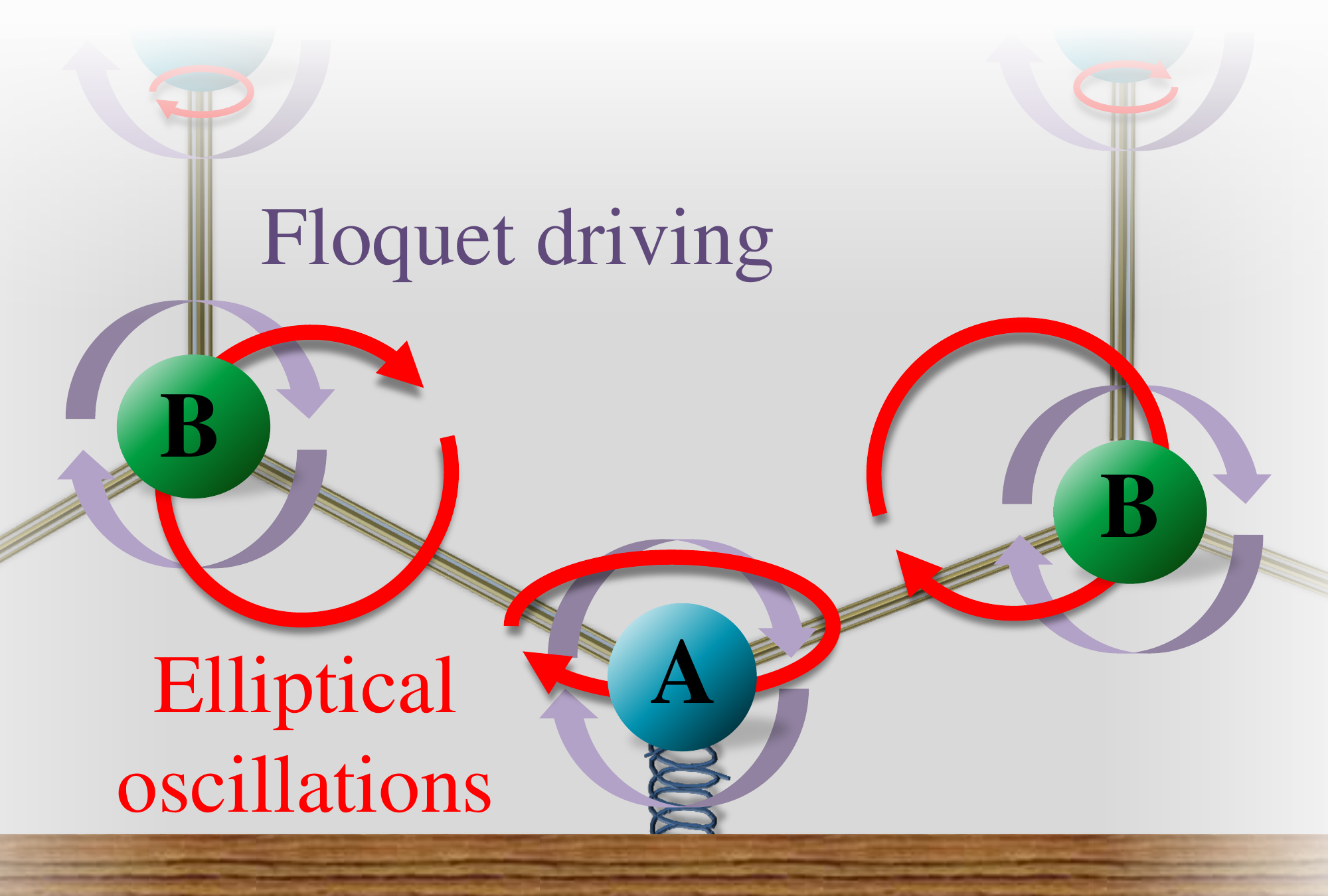}
\includegraphics[width=0.51\linewidth]{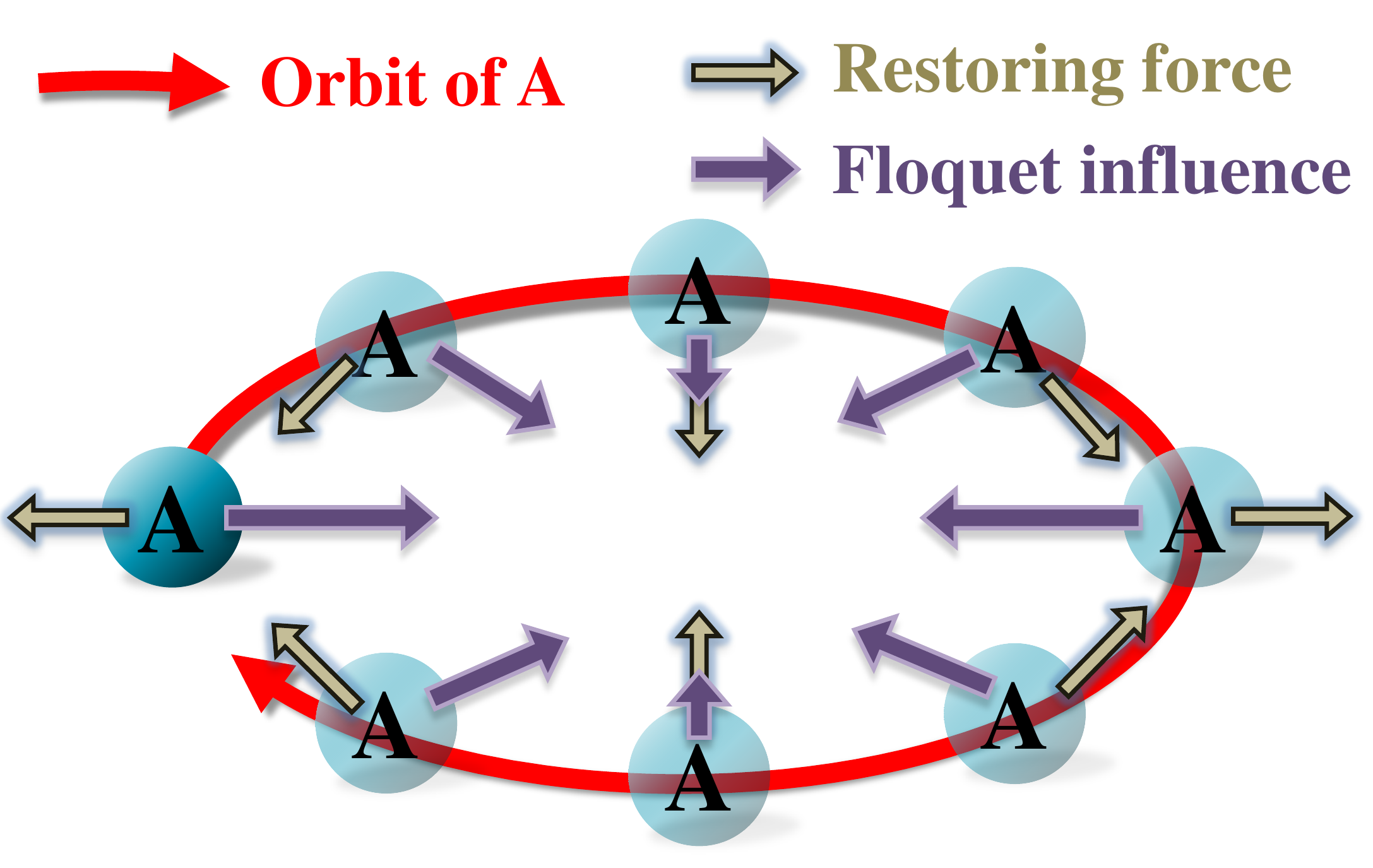}
\end{minipage}
\caption{(Color online) Top Left) Chiral edge modes appear in the Floquet lattice with $\Gamma=1$ and $\Omega=4\sqrt{\frac{k}{m}}$, with the red/blue intensities corresponding to left/right edge localization of amplitude. Top Center) At a sufficiently low $\Omega=3.1\sqrt{\frac{k}{m}}$, inter-BZ chiral modes without static analog appear. Top Right) Simulation of chiral mode propagation in the dynamically modulated system, with initial excitation at the top-left corner. Transient excitations exist within each Floquet period. Bottom) Trajectories of edge masses for the Nyquist topological mode ($k_y=-\frac{\pi}{2}$, $\Omega=2\omega$) with no static analog. Simple balancing of forces reveal the necessity of an onsite dynamical modulation at exactly twice the oscillation frequency.}
\label{fig:floquet}
\end{figure}

Solving for the Floquet dispersion with the time-dependent stiffness matrix given by the non-gyroscopic ($\gamma=0$) honeycomb stiffness matrix plus $K_t(t)$, we obtain plots as shown in Fig. \ref{fig:floquet}. At sufficiently large driving frequency i.e. $\Omega=4\sqrt{\frac{k}{m}}$, the various copies of the Floquet spectra are well-separated and the dispersion resembles that of a static gyroscopic system with nonzero $\gamma$, with bulk bands connected by chiral edge modes of masses ``pushed along'' by the periodic modulation of an otherwise non-chiral system. 

Further insight into the nature of these Floquet chiral modes can be obtained via the Magnus approximation: $H_F^{\text{Mag}}= H_0 + \sum_m \frac1{m\Omega}[H_m,H_{-m}]+O(\Omega^{-2})$. While computations at any finite $\Omega$ will invariably contain inaccuracies, we expect qualitatively correct descriptions of the edge modes of topological phases adiabatically connected to the large $\Omega$ limit. Applying the Magnus approximation to the $\Omega=4\sqrt{\frac{k}{m}}$ case, the effective stiffness matrix (from Eq. \ref{EOM2}) acquires a correction term
\begin{equation}
%\frac1{\Omega}\left(\frac{3\Gamma}{8}\right)^2[\sigma_z-i\sigma_x,\sigma_z+i\sigma_x]\otimes \mathbb{I}
\Delta K_F^{\text{Mag}}=\frac1{\Omega}[K_1,K_{-1}]=-\frac{9\Gamma^2}{16\Omega}\,\sigma_y\otimes\mathbb{I},
\end{equation}
with frequency components given by $K_{\pm 1}=\frac{3\Gamma}{8}(\sigma_z\mp i\sigma_x)\otimes \mathbb{I}$. Comparison with Eq. \ref{EOM} yields an effective frequency-dependent gyroscopic parameter of $|\gamma_F^{\text{Mag}}|=\frac{9\Gamma^2}{16\omega\Omega}$ which agrees well with numerical results of Fig.~\ref{fig:zigzag_large} in the neighborhood of the topological mode crossings. Indeed, detailed examination of the Floquet eigenmodes reveal very similar trajectories with those in Fig. \ref{fig:maze}, whose coincidence is not surprising given the effectively gyroscopic effective Hamiltonian, at least under a narrow range of frequencies. %but instead of velocity-dependent Lorentz forces, we have slightly ...\red{...} 

Notably, our Floquet system also possess ``Nyquist'' topological modes with no static analog. Temporal periodicity folds the quasi-frequency BZ, allowing the the highest and lowest $\omega$ bands to touch and hybridize as the driving frequency $\Omega$ is lowered. In the right quasifrequency plot of Fig.~\ref{fig:floquet}, topological edge modes appear around $k_y = \pm \frac{\pi}{2}$, connecting different copies of the Floquet BZ. These emergent edge modes exist at the Nyquist frequency of $\omega=\Omega/2$, which is ill-defined in static systems. %are the result of the topological phase transition due to the hybridization of the optical bands, and have no static analog. 
It is worth pointing out that the notion of frequency becomes ill-defined when $|\omega|>\Omega/2$, since processes happening below that time-scale have been integrated over. %even though no edge modes connect the optical bulk bands when they are well-separated.
%... In the rightmost panel, the driving frequency is lowered to twice of that of the "optical" bands around $\omega = 1.5\sqrt{\frac{k}{m}}$, showcasing the interplay between the two frequency scales. \red{optical means?} Interestingly, additional edge modes  \red{meaning?}

Explicit solution of these Nyquist topological modes reveal elliptic oscillations that are possible only due to the time-dependent modulation of the stiffness matrix. As illustrated at the Bottom of Fig.~\ref{fig:floquet}, force balance at the minimal at $k_y=\frac{\pi}{2}$, i.e. spatial period of 2 unit cells, requires elliptical trajectories of the edge masses $A$, which are made possible by the fluctuating restoring force that reaches a maximum/minimum when the orbit requires the greatest/least curvature. This exact matching of dynamical modulation with (twice the) oscillation frequency is qualitatively distinct from chiral tendencies due to analogs of the magnetic field.

An experimental setup for the Floquet topological lattice is achievable with basic lab equipment. With an AC current source of amplitude $2.5 A$ connected to $N=400$-loop solenoids of radius $R=2cm$, we obtain magnetic moments $M_0,M_0'=\pi I_0 N R^2=1.26A\cdot m^2$ at their maximum. These moments give rise to effective spring constants amplitudes $\Gamma=\frac{3\mu_0M_0M_0'}{\pi d^5}=6.1N/m$, which can be matched with that of real springs of lengths $a=20cm$ to reproduce Fig. \ref{fig:floquet}. %This is a realistic stiffness for springs of around $a=20cm$ long, the lattice spacing between the solenoids they connect. 
Furthermore, we also require that the electromagnets between different sites interact negligibly. This is easy, since $\Gamma$ falls off as the inverse fifth power of distance. Setting $d=5cm$ (Fig.\ref{fig:floquetexp}), for instance, we find that the electromagnets between neighboring sites couple with a negligible $(5/20)^5\approx 0.1\%$ strength compared to those producing the dynamical driving. Finally, with mobile electromagnets of about $m=230g$, we should expect topological edge modes at frequencies of $\omega\approx \sqrt{\frac{3k}{2m}}\approx 1 Hz$, which are easily observable by an oscilloscope or even the naked eye.

{\sl Conclusions--} By studying gyroscopic and Floquet lattices, we have provided very intuitive illustrations of how topological behavior emerges from Newton's laws. The edge localization and chirality of topological edge modes are shown to result from the ``dangling'' properties of certain boundary sites, a mechanism that holds for generic topological systems governed by 2nd-order ODEs. We have also proposed an experimentally realistic Floquet set-up where the enigmatic behavior of modes with no static analog can be directly observed.
%From a mechanical SSH model, we have built a honeycomb Chern lattice with very easily visualizable edge modes that are also extremely robust. By removing its gyroscopic coupling and introducing dynamical modulation through AC electromagnets, we arrive at a very experimentally realistic Floquet topological Chern lattice whose chiral edge dynamics can be directly observed at macroscopic scales.%\red{add dynamic switching in conclusion if needed}

{\sl Acknowledgements.} Xiao Zhang is supported by the National Natural
Science Foundation of China (No.11404413) and the Natural Science Foundation of Guangdong Province (No.2015A030313188).
\bibliography{references}
\newpage
\onecolumngrid
\begin{center}
\textbf{\large Supplemental Online Material for ``Topological dynamics from Newton's laws'' }\\[5pt]
%Authors\\[5pt]
\vspace{0.1cm}
\begin{quote}
{\small In this supplementary material, we detail: 1) The mechanics of the restoring torque from gyroscopes, 2) The easily visualizable edge mode behavior on a gyroscopic/Lorentz force Chern lattice at arbitrary quasimomentum $p_x$, 3) Our simulation method and 4) Further details on damping effects. }\\[20pt]
\end{quote}
\end{center}
\setcounter{equation}{0}
\setcounter{figure}{0}
\setcounter{table}{0}
\setcounter{page}{1}
\setcounter{section}{0}
\makeatletter
\renewcommand{\theequation}{S\arabic{equation}}
\renewcommand{\thefigure}{S\arabic{figure}}
\renewcommand{\thesection}{S\Roman{section}}
\renewcommand{\thepage}{S\arabic{page}}
\vspace{1cm}
%\twocolumngrid
%\label{SOM}

\section{Equation of motion of masses with gyroscopes attached}
%\label{SOM}
Gyroscopes attached to moving masses provide an experimentally accessible route to the realization of large Lorentz-type forces\cite{nash2015topological} for topological chiral edge modes. Each mass is attached to the free end of a gyroscope with pivot directly above the equilibrium position of the mass. These gyroscopes break time-reversal symmetry by providing a ``reaction torque'' in response to movements of the mass.  %\red{add yuhan's dispersive situation in the SOM as well as a counter example?}

Consider a gyroscope with moment of inertia $I$ and angular spin speed of $ \Psi$ attached to a mass moving with velocity $ \dot{\vec r} =\frac{d\vec r}{dt}$, where the origin of $\vec r$ is taken to be the pivot of the gyroscope. The rate of change of the gyroscopic angular momentum $L=I\Psi\hat r$ is $\frac{d\vec L}{dt}=\frac{d (I\Psi\hat r)}{dt}=I\Psi\dot{\vec r}/|\vec r|$, which must be compensated by a reaction torque $\vec r\times \vec F_{\text{react}}$. Assuming small oscillations so that $\vec r$ is almost perpendicular to the plane of oscillation, we find that 
\begin{equation}
\vec F_{\text{react}}=-i\frac{I\Psi}{h^2}\sigma_2\dot{\vec r}
\end{equation}
where $-i\sigma_2$ implements a $\pi/2$ rotation, and $h$ is the length of the gyroscope. In the space spanned by the phonon polarizations and the sublattices, the equation of motion of the masses hence take the form
\begin{equation}
M\ddot{\vec r}-i\gamma(\sigma_2\otimes \mathbb{I})\dot{\vec r}+K\vec r=0
\label{EOM0}
\end{equation}
where $\vec r$ denote the small displacement of each mass about its equilibrium. Here $M=\mathbb{I}\otimes \text{diag}(m_1,m_2,...)$ is the mass matrix consisting of masses $m_1,m_2,...$ within each unit cell, $K$ is the stiffness matrix, and $\gamma=\frac{I\Psi}{h^2}$ is the gyromagnetic ratio. Note that Eq. \ref{EOM} is of the same form as that of charges $q$ in a magnetic field $B$, if $\gamma$ is replaced by $qB$. In other words, the gyroscopic reaction force behaves exactly like the electromagnetic Lorentz force, at least for small displacements. %However, it is difficult to realize this phonon hall effect with charged masses in a magnetic field, because the electrostatic repulsion between the charges are far stronger (see Appendix A). 

Eq. \ref{EOM0} can be rewritten in first-order form cf. Eq. \ref{EOM2}, which is unitarily equivalent to the alternative phase space formulation of a related problem in Ref. \onlinecite{zhang2010topologicalphonon}. Also, a qualitatively different gyroscopic coupling had been considered in Ref. \onlinecite{wang2015topological}, where the gyroscopic spin depending dynamically with the phonon mode, resulting in a gyroscopic term second order in the time derivative.

\section{S1. Analysis of edge modes of gyroscopic honeycomb lattice }
\label{SOM1}
Consider a gyroscopic honeycomb lattice of identical masses $m$ connected by identical springs with stiffness $k$. Consider the case of a zigzag edge (Fig.\ref{fig:zigzag_large}) with A-type masses sitting on the ``dangling bonds''. Denote the edge direction as being along $\hat x$, so that the momentum $p_x$ is still a good quantum number. An edge eigenmode is characterized by a decay factor $|t|=e^{-1/L}<1$, such that the amplitudes of its oscillations decay exponentially perpendicularly from the edge like $|t|^y=e^{-y/L}$.
 
As explained in the main text, the edge modes of a honeycomb lattice consists of A-type masses moving around in small ellipses, and B-type masses oscillating entirely along $\hat x$, reminiscent of the stationary B-type masses in the SSH chain. We expand an edge mode in terms of plane waves indexed by momenta $p_x$, with phase velocity given by $\omega/p_x$.  

For each unit cell $n$ along the $\hat x$ direction, denote the displacements of masses $A$ and $B$ from their equilibria due to a $p_x$-eigenmode by 
\begin{subequations}
\begin{equation}
\vec r_{A,n}=(x_{A,n},y_{A,n})^T=(x_A\sin(p_xn-\omega t),~ y_A\cos(p_xn-\omega t))^T 
\end{equation}
\begin{equation}
\vec r_{B,n}=(x_{B,n},y_{B,n})^T=(-x_B\sin(p_xn-\omega t),~0)^T 
\end{equation}
\end{subequations}
i.e. the $A$ atoms make small clockwise elliptical oscillations while $B$ atoms strictly vibrate in the $\hat x$ plane. The honeycomb lattice consists of $A$ and $B$ sites displaced by half a unit cell spacing in the $\hat x$ direction. In addition to the spring restoring forces, each mass experiences a Lorentz force of $\gamma ~  \dot \vec r\times \hat z$, where $\gamma$ is the gyroscopic coupling. By balancing the accelerations and forces on masses $A$ and $B$ in the $\hat x$ and $\hat y$ directions respectively, we obtain
\begin{align}
m\ddot x_{A,n}&=\gamma\dot y_{A,n}+\frac{3k}{4}\left[(x_{B,n-1/2}-x_{A,n})+(x_{B,n+1/2}-x_{A,n})\right]\\
m\ddot y_{A,n}&=-\gamma \dot x_{A,n}-\frac{3k}{2}y_{A,n}-\frac{\sqrt{3}k}{4}\left[(x_{B,n-1/2}-x_{A,n})-(x_{B,n+1/2}-x_{A,n})\right]\\
m\ddot x_{B,n+1/2}&=\frac{3k}{4}\left[(x_{A,n}-x_{B,n+1/2})+(x_{A,n+1}-x_{B,n+1/2})\right]+\frac{\sqrt{3}k}{4}\left[(y_{A,n}-0)-(y_{A,n+1}-0)\right]\\
\gamma \dot x_{B,n+1/2} &=\frac{k}{2}\left[(y_{A,n}-0)+(y_{A,n+1}-0)\right] +k[ty_{A,n+1/2}]+\frac{\sqrt{3}k}{4}\left[(x_{A,n}-x_{B,n+1/2})-(x_{A,n+1}-x_{B,n+1/2})\right]
\end{align}
Since it was assumed that masses of type $B$ do not move in the $\hat y$ direction i.e. $y_B=0$, there is no Lorentz force contribution to $m\ddot x_B$ in line 3. This further implies that in the $\hat y$ direction, the Lorentz force exactly cancels the spring restoring forces on a mass of type $B$ (line 4). Each of such a mass couples to exactly one $A$-type mass situated an unit cell further from the edge, whose oscillations are attentuated by a factor of $\pm t$, the $\pm$ sign allowing for a possible reversal in oscillation direction. Simplifying the above, we obtain
%\begin{subequations}
\begin{eqnarray}
\label{edgeeq1}
(\omega^2-\omega^2_0)\eta_A&=&-\Gamma \omega  +\omega^2_0\eta_B\cos \frac{p_x}{2}\\
\omega^2-\omega^2_0&=&-\Gamma \omega \eta_A -\frac{\omega^2_0}{\sqrt{3}}\eta_B\sin \frac{p_x}{2}\\
(\omega^2-\omega^2_0)\eta_B&=&\omega^2_0\eta_A\cos \frac{p_x}{2}-\frac{\omega^2_0}{\sqrt{3}}\sin\frac{p_x}{2}\label{edgeeq3}\\
\Gamma \omega \eta_B&=& \frac{\omega^2_0}{3}\left(\cos \frac{p_x}{2}\pm 2t\right)+\frac{\omega^2_0}{\sqrt{3}}\eta_A\sin \frac{p_x}{2}
\label{edgeeq4}
\end{eqnarray}
where $\eta_A=\frac{x_A}{y_A}$, $\eta_B=\frac{x_B}{y_A}$, $\Gamma=\frac{\gamma }{m}$ is the normalized gyroscopic coupling and $\omega_0=\sqrt{\frac{3k}{2m}}$ denotes the resonant frequency of each tri-bond. $|\eta_A|$ denotes the aspect ratio of mass A's orbit, with positive/negative $\eta_A$ indicating anticlockwise/clockwise oscillations.

Eqs. \ref{edgeeq1} to \ref{edgeeq4} can be simultaneously solved to yield $\omega$, $\eta_A$, $\eta_B$ and $L$ in term of $p_x$ and $\Gamma$. In particular, $\omega(p_x)$ agrees exactly with the numerical dispersion of the edge modes displayed in Fig.\ref{fig:zigzag_large}. 

\subsection{Long wavelength ($p_x=0$) limit}
It is instructive to %first
study the spatially homogeneous edge modes, i.e. those with $p_x=0$ (at the Gamma-point). In this limiting case Eqs. \ref{edgeeq1} to \ref{edgeeq4} reduce to 
\begin{eqnarray}
\label{edgeeq5}
(\omega^2-\omega^2_0)\eta_A&=&-\Gamma \omega  +\omega^2_0\eta_B\\
\label{edgeeq6}
\omega^2-\omega^2_0&=&-\Gamma \omega \eta_A  \\
\label{edgeeq7}
(\omega^2-\omega^2_0)\eta_B&=&\omega^2_0\eta_A\\
\Gamma \omega \eta_B&=& \frac{\omega^2_0}{3}\left(1\pm 2t\right)
\label{edgeeq8}
\end{eqnarray}
In particular, Eqs. \ref{edgeeq5} to \ref{edgeeq7} can be combined to yield $(\omega^2-\omega_0^2)(\eta_A^2-\eta_B^2-1)=0$, with an obvious solution $\omega=\omega_0=\sqrt{\frac{3k}{2m}}$, $\eta_A\propto x_A=0$.  This corresponds to type-A and B masses oscillating orthogonally, along the $\hat y$ and $\hat x$ axes respectively. From Eq. \ref{edgeeq5}, their relative oscillation amplitudes are given by 
\begin{equation}
\frac{x_B}{y_A}=\eta_B=\frac{\gamma}{m\omega}=\sqrt{\frac{2\gamma^2}{3mk}}
\label{eqB}
\end{equation}
In other words, with nonzero gyroscopic coupling $\gamma$ (or magnetic force), mass $B$ has to oscillate horizontally to compensate the Lorentz force on mass $A$. This can also be easily seen by equating the Lorentz force on the latter, which is proportional to $\gamma \omega y_A$, with the $\hat x$-direction inertia force from $b$, which is proportional to $m\omega^2 x_B$.
 
Together with Eq. \ref{edgeeq8}, the result Eq. \ref{eqB} gives the edge mode decay length as 
\begin{equation}
L=-(\log|t|)^{-1}=-\left [\log\frac{|3\eta^2_B-1|}{2}\right]^{-1}=-\left [\log\left|\frac{\gamma^2}{mk}-\frac1{2}\right|\right]^{-1}
\end{equation}
A topological phase transition occurs when the gap closes and $L$ diverges, which occurs at $\left|\frac{2\gamma^2}{mk}-1\right|=|3\eta_B^2-1|=2$. In particular, this means that the gap around $\omega=\omega_0$ becomes topologically trivial at sufficiently large gyroscopic coupling $\gamma$ i.e. $\gamma>\sqrt{\frac{3mk}{2}}$, when the Lorentz force is too strong to be compensated by an oscillatory mode decaying into the bulk (Eq. \ref{edgeeq8}). Such a phase transition arises due to the interplay between the magnetic force and the lattice springs, and does not exist in a continuum quantum Hall system.

Interestingly, it is also possible to tune $\gamma$ such that the edge mode is perfectly localized at the edge. At the special value $\gamma=\sqrt{\frac{mk}{2}}$, $L=0$ and there is \emph{totally no vibration} beyond the boundary $A$ and $B$ sites. This occurs when the Lorentz force on mass $B$ can be completely compensated by the restoring motion of the edgemost $A$ masses alone.

%with positive coupling constant forces $\eta_A=0$, i.e. that $A$-type atoms vibrate entirely in the $h$
\begin{figure}%[H]
\begin{minipage}{\linewidth}
\includegraphics[width=0.24\linewidth]{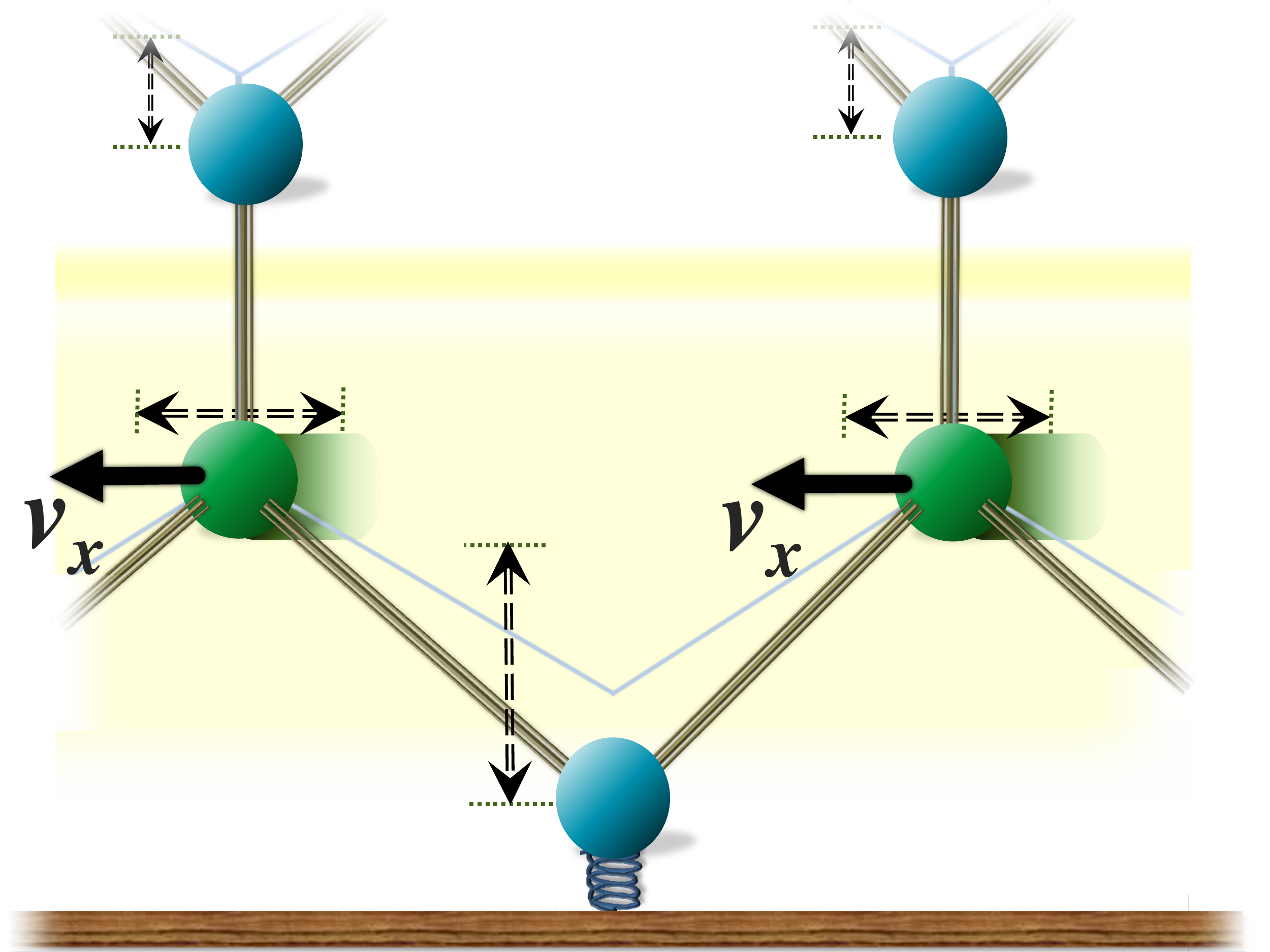}
\includegraphics[width=0.24\linewidth]{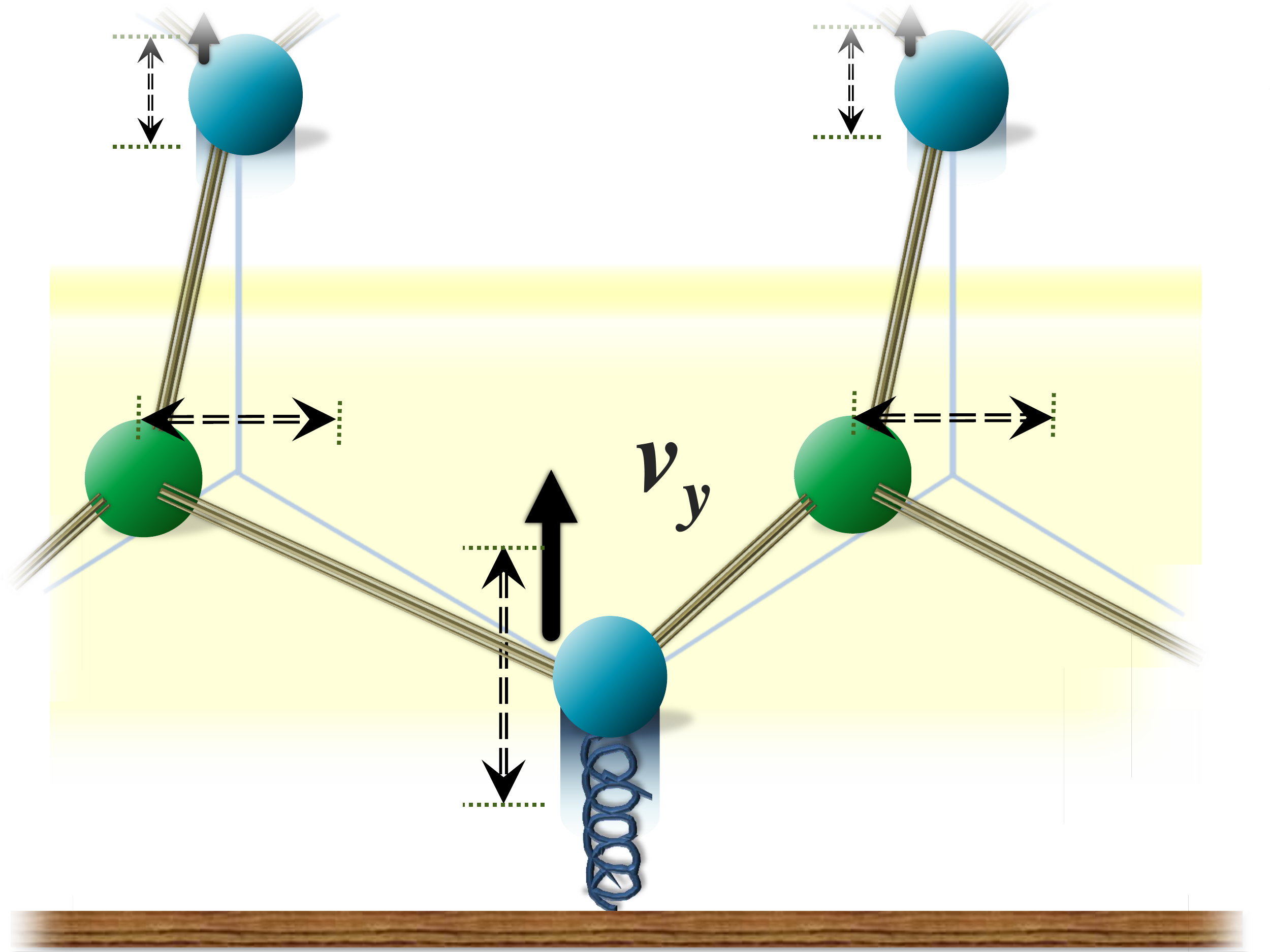}
\includegraphics[width=0.24\linewidth]{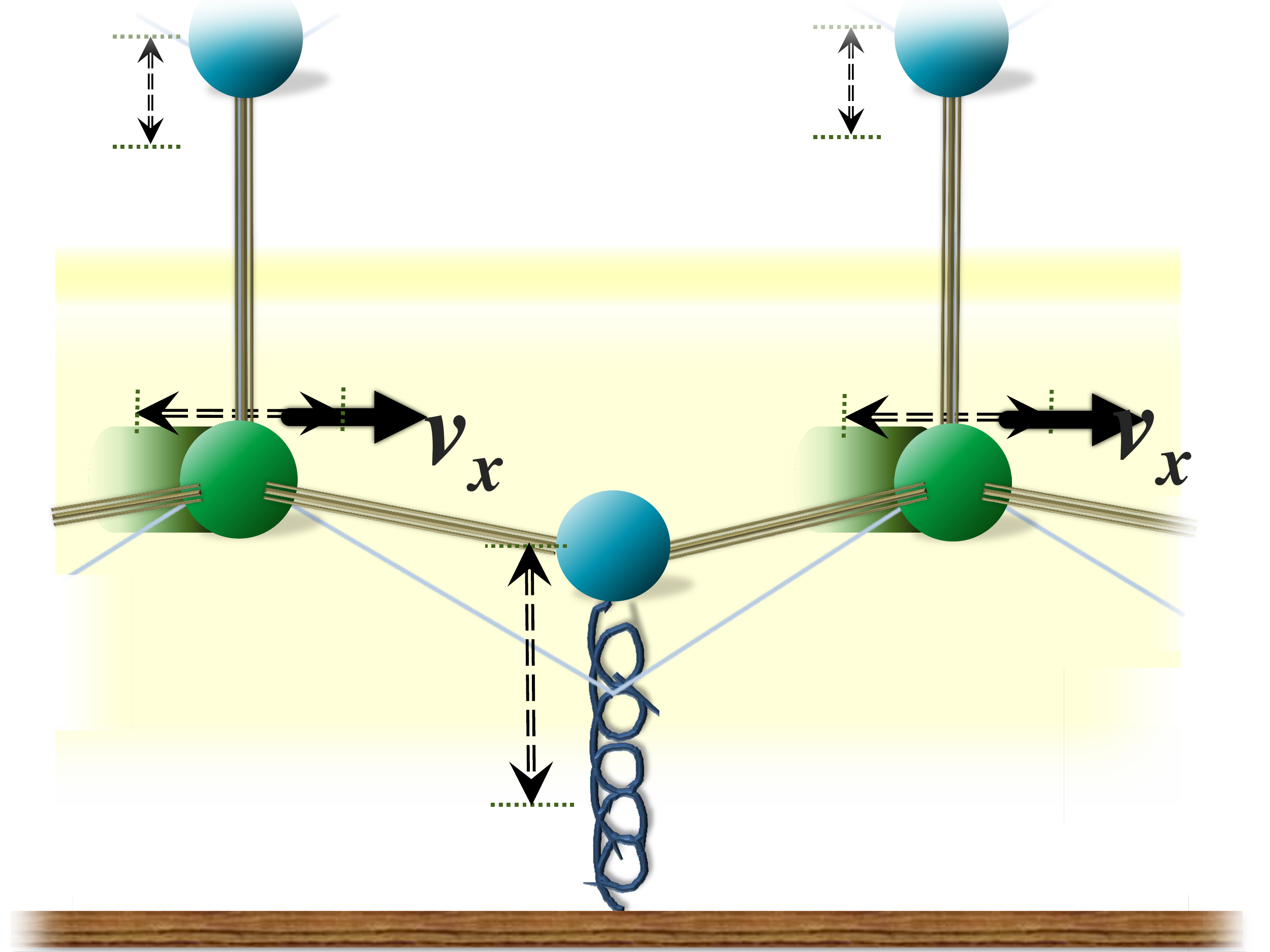}
\includegraphics[width=0.24\linewidth]{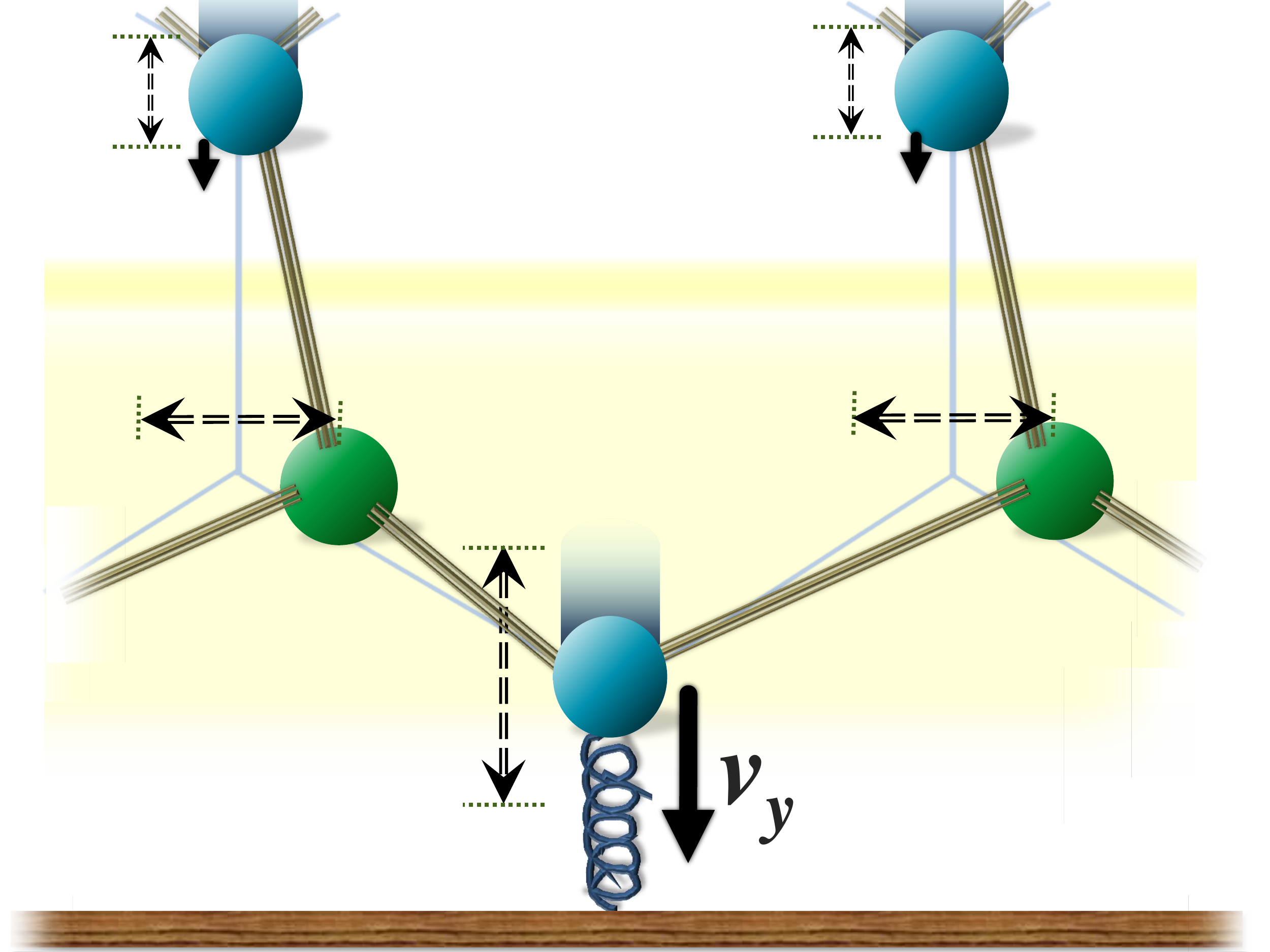}
\end{minipage}
\caption{(Color online) Snapshots at every quarter cycle of the chiral edge mode oscillation at the $\Gamma$ point ($p_x=0$). The Lorentz force on each mass cancels the exactly synchronized restoring forces from the neighboring masses, thereby maintain strictly vertical/horizontal motion for the A/B-type masses.}
\label{fig:hc14}
\end{figure}

\subsection{Edge modes at generic $p_x$}

For generic quasi-wavevector $p_x$, the dispersion of the edge modes can be obtained by simultaneously solving Eqs. \ref{edgeeq1} to \ref{edgeeq3}. The resultant solution can be substituted into Eq. \ref{edgeeq4}, from which $t$ can be obtained. The mode merges into the bulk when $t\geq 1$. After some simplification, $\omega$ is given by the solution of
\begin{equation}
(1-\cos p_x)-2(3+\Gamma^2)W^2+3W^4=\frac{\sqrt{2}W \sin p_x}{W^2-1}
\end{equation}
where $W=\omega/\omega_0$, $\omega_0=\sqrt{\frac{3k}{2m}}$. Numerically, these solutions can be shown to coincide exactly with the edge modes (and some bulk modes) of the phonon dispersions diagrams in Fig. 1 of the main text.

The case with zero gyroscopic coupling ($\Gamma=0$) admits a full analytical solution. The three positive frequency modes are:
\begin{itemize}
\item the flat dispersion mode $\omega=\sqrt{\frac{3}{2}}$ with localization lengths $L=(\log\left|2\cos\frac{p_X}{2}\right|)^{-1}$ and aspect ratios of mass A's orbit being $\eta_A=\frac1{\sqrt{3}}\tan\frac{p_x}{2}$. Evidently, these modes merge into the bulk ($L$ becomes negative) when $|p_x|>\frac{2\pi}{3}$, as presented numerically in Fig. 1. When this occurs, mass $A$'s orbit is exactly circular. 
\item the pair of dispersive modes $\omega=\sqrt{\frac{3\pm\sqrt{3(2+\cos p_x)}}{2}}$ with localization lengths $L=-(\log\left|\cos\frac{p_X}{2}\right|)^{-1}$ and aspect ratios of mass A's orbit being $\eta_A=-\sqrt{3}\cot\frac{p_x}{2}$. In the long-wavelength limit, $L\sim \frac{8}{p_x^2}$ diverges quadratically, with the orbit of mass $A$ becoming infinitely elongated.  
%\red{please give some comments, or indicate what else is interesting, for me to solve}
\end{itemize}

\section{Simulation method}

We performed numerical simulations for the lattice dynamics shown in Figs. 2 and 4 of the main text. The force equation (Eqs. 1 and 9 of the main text) was solved via an ODE solver (ode45 of MATLAB R2014b) employing the a Fourth-order Runge-Kutta algorithm. For maximum accuracy, we used a timestep of $0.005~s$ in units of $\sqrt{\frac{m}{k}}$, such that each oscillation corresponds to a few hundred timesteps. The illustrative systems we solved contain $900$ sites in the A sublattice and $841$ sites in B sublattice, and correspond to a $6964\times 200001$ matrix for the dynamical motion. 

Within each simulation, a chosen mass (Top Right corner for the gyroscopic lattice and Top Left corner for the Floquet lattice) is constantly driven at frequency $\omega$. This mass acts as the driving source of lattice vibrations, and should be distinguished from Floquet driving which is a simultaneous time modulation of the stiffness matrix on \emph{all} sites, at frequency $\Omega$.

The detailed trajectories of all other masses is tracked for a duration of $1000~s$ ($2\times 10^5$ timesteps), and will be displayed in the supplementary .gif files. The initial 20 seconds trajectories of the first few neighbors of the driving source are also presented and discussed in Fig 2. 

\section{Effect of damping}
We studied how damping affects the lifetimes and hence propagation lengths of the topological edge modes in both the gyroscopic and Floquet lattices. For the former, the equation of motion is modified to
\begin{equation}
M\ddot{\vec r}+\beta(\mathbb{I}\otimes \mathbb{I})\dot{\vec r}-i\gamma(\sigma_2\otimes \mathbb{I})\dot{\vec r}+K\vec r=0
\label{EOMdamped}
\end{equation}
where the damping parameter $\beta$ is assumed to be the same for all sites. The Floquet system has $\gamma=0$, but with $K\rightarrow K(t)$ taking on a time dependence described in Eq. 9 of the main text. As presented in Fig. 2, there is a clear inverse relationship between $\beta$, the amount of damping and $\xi$, the propagation length of an edge mode from its driving source.

\begin{figure}%[H]
\begin{minipage}{\linewidth}
\includegraphics[width=0.64\linewidth]{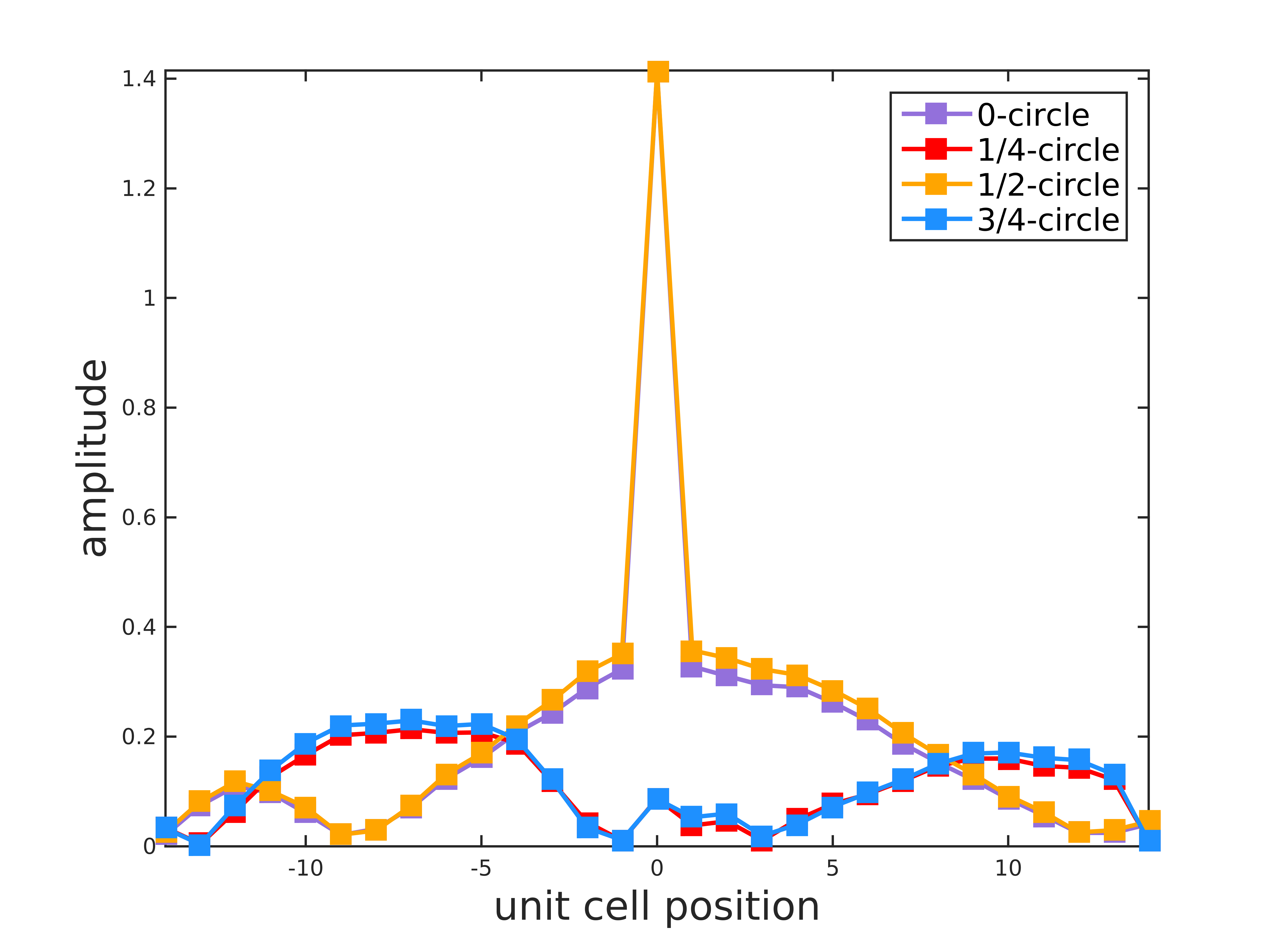}
\end{minipage}
\caption{(Color online) The spatial displacements within an illustrative Floquet edge mode within 15 unit cells away from the driving source along an edge. Shown are the amplitudes of each edge mass after completing $\frac1{4},\frac1{2},\frac{3}{4}$ or all of a full oscillation. Averaging out these phases, we obtain an approximately exponential spatial decay profile.
}
%\label{fig:damping}
\end{figure}

%\section{Gryo...}

%\begin{figure}%[H]
%\begin{minipage}{\linewidth}
%\includegraphics[width=0.32\linewidth]{singlegyroscope.pdf}
%\includegraphics[width=0.32\linewidth]{hexgyro3D.pdf}
%\includegraphics[width=0.32\linewidth]{checkerboard3D.pdf}
%\end{minipage}
%\caption{(Color online) add bandstructures....}
%\label{fig:gyro}
%\end{figure}

%\end{subequations}

\end{document}